\documentclass[twocolumn,english,aps,prd,graphics,floatfix,a4paper,superscriptaddress]{revtex4-1} % {revtex4}
\usepackage{amsmath, amssymb}
\usepackage{makecell}
\usepackage{multirow}
\usepackage{CJK}
\usepackage{graphicx}
\usepackage{mathrsfs}
\usepackage{bm}
\usepackage{amsmath}
\usepackage{dcolumn}
\usepackage{epstopdf}
\usepackage{dsfont}
\usepackage{amssymb}
\usepackage{tabularx}
\usepackage{array}
\usepackage{float}
\usepackage{color}
\usepackage{epstopdf}
\usepackage{mathrsfs}

\usepackage[colorlinks, linkcolor=blue,anchorcolor=blue,citecolor=blue,urlcolor=blue]{hyperref}
\usepackage{extarrows}

\begin{document}
	
\title{Electric Edelstein Effect in Two Dimensions}

\author{Xiaoran Lv}\thanks{These authors contribute equally}
\affiliation{Key Laboratory of Low-Dimensional Quantum Structures and Quantum Control of Ministry of Education, Department of Physics and Synergetic Innovation Center for Quantum Effects and Applications, Hunan Normal University, Changsha 410081, China}

\author{Tianqi Zhang}\thanks{These authors contribute equally}
\affiliation{Key Laboratory for Liquid-Solid Structural Evolution and Processing of Materials, Ministry of Education, Shandong University, Jinan 250061, China}

\author{Chaoxi Cui}\email{cuichaoxi@gmail.com}
\affiliation{Key Laboratory of Advanced Optoelectronic Quantum Architecture and Measurement (MOE),
School of Physics, Beijing Institute of Technology, Beijing 100081, China}

\author{Yaling Li}
\affiliation{Key Laboratory of Low-Dimensional Quantum Structures and Quantum Control of Ministry of Education, Department of Physics and Synergetic Innovation Center for Quantum Effects and Applications, Hunan Normal University, Changsha 410081, China}

\author{Yi Shen}
\affiliation{Key Laboratory of Advanced Optoelectronic Quantum Architecture and Measurement (MOE),
School of Physics, Beijing Institute of Technology, Beijing 100081, China}

\author{Weikang Wu}\email{weikang\_wu@sdu.edu.cn}
\affiliation{Key Laboratory for Liquid-Solid Structural Evolution and Processing of Materials, Ministry of Education, Shandong University, Jinan 250061, China}

\author{Ziming Zhu}\email{zimingzhu@hunnu.edu.cn}
\affiliation{Key Laboratory of Low-Dimensional Quantum Structures and Quantum Control of Ministry of Education, Department of Physics and Synergetic Innovation Center for Quantum Effects and Applications, Hunan Normal University, Changsha 410081, China}

\author{Yugui Yao}%\email{ygyao@bit.edu.cn}
\affiliation{Key Laboratory of Advanced Optoelectronic Quantum Architecture and Measurement (MOE),
School of Physics, Beijing Institute of Technology, Beijing 100081, China}

\author{Zhi-Ming Yu}%\email{zhiming\_yu@bit.edu.cn}
\affiliation{Key Laboratory of Advanced Optoelectronic Quantum Architecture and Measurement (MOE),
School of Physics, Beijing Institute of Technology, Beijing 100081, China}

\begin{abstract}

We propose the electric Edelstein effect (EEE) in two-dimensional (2D) systems, where crossed out-of-plane and in-plane electric fields generate a spin polarization. The gate field converts the Edelstein response from a material-fixed spin polarization into a gate-tunable one. A symmetry analysis over all magnetic layer point groups shows that the EEE is allowed in a wide range of magnetic and nonmagnetic systems, including ferromagnets, antiferromagnets, and altermagnets. The EEE plays two distinct roles: it generates spin polarization where the linear Edelstein effect (EE) is symmetry-forbidden, as in centrosymmetric materials; and it adds a new gate-tunable component, generally non-collinear with the conventional one, where the EE is already present. The two fields then act as independent knobs and raise the control dimension of the spin polarization from one to two. Our microscopic theory traces the EEE to band-geometric quantities, which are strongly enhanced near small band gaps. First-principles calculations predict pronounced effects in the centrosymmetric ferromagnet MnSe$_2$, where the linear EE is symmetry-forbidden. In the nonmagnet SnSe where the EE and EEE coexist, the two knobs cover an extended region of the spin sphere. Our findings establish the EEE as a symmetry-guided, gate-tunable route to electrical spin manipulation and open avenues for multistate spintronics in 2D quantum materials.
\end{abstract}

\maketitle
\begin{figure}[t]
			\centering
			\includegraphics[width=0.5\textwidth]{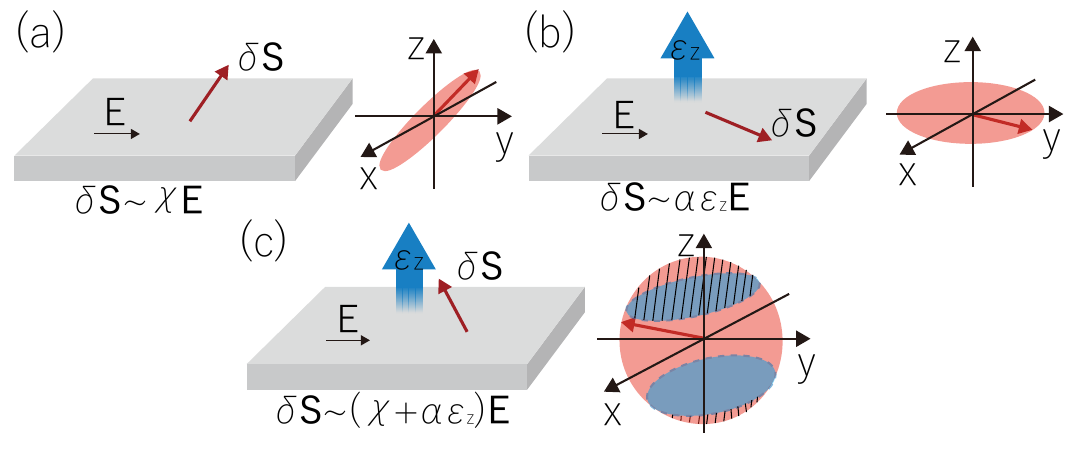}
			\caption{{(a) and (b) show that the spin polarizations from the conventional EE and pure EEE are confined to the 2D-plane, respectively. (c) shows the rotation direction of spin polarization from the coexistence of conventional EE and EEE can reach a large portion of spin-sphere.}}
			\label{rotate}
		\end{figure}

%\end{CJK*}
{\color{blue}\textit{Introduction.---}}
The generation and manipulation of spin by purely electrical means remains a central challenge in spintronics. The Edelstein effect (EE), or the current-induced spin polarization, is a key mechanism: an in-plane charge current generates a net spin polarization (see Fig.~\ref{rotate}(a)) in materials without spatial inversion symmetry ($\mathcal{P}$)  \cite{manchon2019current,awschalom2009spintronics,aronov1989nuclear,edelstein1990spin,fang2011spin,ivchenko1978new,kato2004current,zelezny2014relativistic,zhou2026layer}, and also in heterostructures \cite{mihai2010current,miron2011perpendicular,liu2012spin}. Under time reversal ($\mathcal{T}$), the EE decomposes into $\mathcal{T}$-even and $\mathcal{T}$-odd contributions \cite{manchon2019current,freimuth2014spin,vzelezny2017spin,ivchenko1978new,aronov1989nuclear,edelstein1990spin,culcer2007generation}. The former exists in both magnetic and nonmagnetic materials, while the latter appears only in magnets. These two components generate two fundamental classes of spin-orbit torques and thereby enable the electrical control of magnetic order \cite{manchon2019current,aronov1989nuclear,edelstein1990spin,xiao2022intrinsic,xiao2023time}.

%A cornerstone in the field is the Edelstein effect (EE), or current-induced spin polarization, in which an in-plane charge current generates a net spin polarization in materials lacking spatial inversion symmetry ($\mathcal{P}$), as well as in heterostructures. Depending on its behavior under time-reversal ($\mathcal{T}$) symmetry, the EE decomposes into $\mathcal{T}$-even and $\mathcal{T}$-odd contributions: the former exists in both magnetic and nonmagnetic materials, while the latter is exclusive to magnetic systems. These two components generate two fundamental classes of spin-orbit torques and thereby enable the electrical control of magnetic order.

The conventional EE is limited in two respects. First, it is symmetry-restricted. Because $\mathcal{P}$ must be broken, a wide class of centrosymmetric two-dimensional (2D) materials has no linear EE. Second, it offers limited tunability. The response is linear in the applied field, $\delta s_a=\chi_{ab}E_b$, so the induced spin polarization is fixed by the material-specific tensor $\chi_{ab}$. Sweeping the in-plane field $\bm{E}$ therefore merely rotates $\delta\bm{s}$ within a fixed plane in spin space, and tilting it out of that plane requires a second, independent knob.
%Second, it offers limited tunability. Being linear in the applied field, i.e. $\delta s_a=\chi_{ab}E_b$, the induced spin polarization is fixed by the material-specific tensor $\chi_{ab}$. Since $\delta\bm{s}$ is the image of a single in-plane vector $\bm{E}$ under this fixed map, its direction is confined to one rotation plane in spin space, and no independent knob can tilt it out of that plane.

Recent efforts to circumvent the symmetry restriction have turned to a nonlinear paradigm \cite{xiao2022intrinsic,xiao2023time,xu2021light,xu2025linear,sarkar2025light,feng2025quantum,cao2026electrical,xiang2025spin,jia2026nonlinear}, in which the induced spin magnetization scales quadratically with the applied electric field. For example, the second-order EE can arise in centrosymmetric materials, driven by the momentum-space dipole of the anomalous spin polarizability (ASP) \cite{xiao2023time}; the nonlinear magnetoelectric EE instead combines electric and magnetic fields, with spin-space Berry-curvature dipoles and quantum-metric dipoles mediating the response \cite{jia2026nonlinear}. Both routes restore the EE in centrosymmetric systems, but neither addresses its tunability: the induced spin polarization remains fixed by the material-specific properties and confined to a single plane in spin space.

%Recent efforts to circumvent the symmetry restriction have turned to a nonlinear paradigm, in which the induced spin magnetization scales quadratically with the applied electric field. For instance, the second-order EE can appear in centrosymmetric materials, related to the momentum-space dipole of the anomalous spin polarizability (ASP). The nonlinear magnetoelectric EE generates spin magnetization from the cooperative action of electric and magnetic fields via spin-space Berry-curvature dipoles and quantum-metric dipoles. These nonlinear routes restore the EE in centrosymmetric systems, yet the induced spin polarization remains fixed by the material and confined to the plane, so the tunability limitation remains unaddressed.

Among the external fields available in 2D systems, the vertical electric field $\mathcal{E}_z$ stands out, which addresses both limitations at once. Because the electrons are confined to the 2D plane, $\mathcal{E}_z$ drives no in-plane current and acts instead as a highly tunable gate that modulates the electronic structure \cite{cui2025electric,sahoo2005electric,rovillain2010electric,heron2014deterministic,gao2020tunable,chen2024crossed,fan2024intrinsic,li2024dynamical,ohno2000electric,fert2024electrical,manchon2019current,liu2012spin}. First, $\mathcal{E}_z$ breaks the spatial symmetries that forbid the EE, such as $\mathcal{P}$ and $M_z$, and thus enables spin generation in a broad class of centrosymmetric 2D materials where the linear EE is symmetry-forbidden. Second, it supplies an independent control knob, so the induced spin polarization is no longer fixed by the material alone. Together, these two roles make $\mathcal{E}_z$ a natural route to electrical spin generation and control in 2D. Yet this route remains largely unexplored and the band-geometric origin for the $\mathcal{E}_z$-induced EE is still unclear.

In this letter, we propose the electric Edelstein effect (EEE) where the spin polarization is generated by the interplay of crossed out-of-plane and in-plane electric fields (see Fig.~\ref{rotate}(b)). The EEE upgrades the spin polarization from a material-fixed, planar quantity to a gate-tunable one, and it does so in two complementary ways. First, it generates spin polarization in systems where the linear EE is symmetry-forbidden. Second, it adds a gate-tunable component in systems that already host the EE. In the latter case, the in-plane field and the gate field act as two independent knobs, which raise the control dimension of the spin polarization from one to two (see Fig.~\ref{rotate}(c)). We establish the symmetry conditions of the EEE over all magnetic layer point groups (MLPGs) \cite{wang2026type}, derive its microscopic expression in terms of band-geometric quantities, and demonstrate both scenarios in realistic materials with first-principles calculations.

In each scenario, the induced spin polarization is composed of two contributions that are in general not collinear. When the linear EE is symmetry-forbidden, the EEE acts alone and can split into mutually orthogonal $\mathcal{T}$-even and $\mathcal{T}$-odd components \cite{manchon2019current,xiao2022intrinsic,xiao2023time}, as in the centrosymmetric ferromagnet MnSe$_2$. More intriguingly, when the EE is already present, the EEE coexists with it as an independent contribution. In suitable materials, the two contributions are orthogonal, and sweeping the gate field would rotate the spin polarization from a single plane to one that covers a large portion of the spin sphere. The nonmagnetic 2D material SnSe is a promising candidate. This provides a principled route to spin control over an extended region of the spin sphere through the Edelstein effect, and establishes the EEE as a symmetry-guided, gate-tunable mechanism for electrical spin generation and control in 2D quantum materials.

{\color{blue}\textit{Symmetry conditions for EEE.}}
The EEE is described by the following response relation:
\begin{equation}
    \delta s_a = \alpha_{ab} \mathcal{E}_z E_b,
\end{equation}
where $\alpha$  is the EEE coefficient, $\delta s_a$ is the induced spin polarization, and $E_b$ is the applied in-plane electric field. The roman indices label the Cartesian components, i.e., $a \in \{x, y, z\}$ and $b \in \{x, y\}$.  The Einstein summation convention is adopted.

To identify the symmetry-allowed elements of $\alpha$, we carry out a symmetry analysis of 2D crystals, constrained by MLPGs for 2D materials \cite{wang2026type}. The gate field is odd under any operation that reverses the out-of-plane \textit{z} direction, such as $\mathcal{P}$, the horizontal mirror \textit{${M}_z$}, and the in-plane rotation ${C}_{2x/y}$. The additional sign from $\mathcal{E}_z$ makes these operations constrain the EEE differently from the linear EE. For example, in centrosymmetric 2D materials, $\mathcal{P}$ forbids the linear EE entirely but places no constraint on the EEE.

% Although the EE is forbidden for some 2D materials, $\mathcal{E}_z$ can break any operator that contains \textit{z} transformation, including $\mathcal{P}$, horizontal mirror \textit{${M}_z$}, in-plane rotation ${C}_{2x/y}$ and so on, and thus may induce EEE. The EEE can be described by the following response relation:
% \begin{equation}
%     \delta s_a = \alpha_{ab} \mathcal{E}_z E_b,
% \end{equation}
% where $\alpha$  is the EEE coefficient, $\delta s_a$ is the induced spin polarization, and $E_b$ is the applied in-plane electric field. The roman indices label the Cartesian components, i.e., $a \in \{x, y, z\}$ and $b \in \{x, y\}$.  The Einstein summation convention is adopted.

The EEE coefficient $\alpha$ can be uniquely decomposed into a $\mathcal{T}$-even part and a $\mathcal{T}$-odd part, $\alpha = \alpha_{\text{even}} + \alpha_{\text{odd}}$. $\alpha_{\text{even}} (\alpha_{\text{odd}})$ is an odd (even) function of the relaxation time $\tau$, which represents the extrinsic (intrinsic) contribution to $\mathcal{E}_z$-induced spin polarization. The two parts obey different symmetry rules. Under an  operation $\mathcal{O}$ of an MLPG, $\alpha$ is constrained by
\begin{equation}\label{eq:alpha_transform}
    \alpha_{a'b'} = \eta_{\mathcal{T}} \eta_z \det(\mathcal{O})\mathcal{O}_{a'a} \mathcal{O}_{b'b} \alpha_{ab},
\end{equation}
where $\det(\mathcal{O})$ is the determinant of the matrix representation of $\mathcal{O}$, $\eta_{\mathcal{T}}=\pm 1$ denote the transformation properties of the system under $\mathcal{T}$, and $\eta_z = +1$ $(-1)$ if $\mathcal{O}$ preserves (flips) \textit{z}. For $\alpha_{\text{even}}$, $\eta_{\mathcal{T}} = +1$ regardless of $\mathcal{O} \equiv R$ ($R$ denotes a unitary operator) or $\mathcal{O} \equiv R\mathcal{T}$, while for $\alpha_{\text{odd}}$, $\eta_{\mathcal{T}} = +1$ when $\mathcal{O} \equiv R$ and $\eta_\mathcal{T} = -1$ when $\mathcal{O} \equiv R\mathcal{T}$.

% The $\mathcal{T}$-even and $\mathcal{T}$-odd components of the EEE is summarized in Table \ref{tab:even} and Table~\ref{tab:odd}. 
% Our results show that the $\mathcal{T}$-even ($\mathcal{T}$-odd) EEE transforms as a third-rank pseudotensor and is theoretically supported by ... MLPGs. In nonmagnetic materials, only the $\mathcal{T}$-even EEE exists, while the $\mathcal{T}$-odd EEE appears only in magnetic materials. Our analysis further reveals that the direction of $\delta\bm s$ is determined by the crystalline point-group symmetries. For instance, in systems with $C_{nz}$ symmetry for $n \ge 3$, the in-plane components of $\delta \mathbf{s}$ must be  isotropic ($\alpha_{xx}= \alpha_{yy}$), while $M_z$ or $C_{nz}$ forbids $\delta s_z$.

% Overall, the EEE alone is enabled in 70 (25) MLPGs for the $\mathcal{T}$-even ($\mathcal{T}$-odd) response. The mixed spin polarization, in while the conventional EE and the EEE coexist, is supported by 78 magnetic point groups and 22 nonmagnetic point groups (see Supplemental Material).

The $\mathcal{T}$-odd and $\mathcal{T}$-even components of the EEE are summarized in Tables \ref{tab:even} and \ref{tab:odd}. Both transform as third-rank pseudotensors. The $\mathcal{T}$-odd component requires magnetic order, so it vanishes in nonmagnetic materials where only the $\mathcal{T}$-even component survives. Our analysis further reveals that the crystalline point-group symmetry fixes the direction of the induced spin polarization. For instance, in systems with $C_{nz}$ symmetry for $n \ge 3$, the in-plane response of $\delta \mathbf{s}$ must be isotropic ($\alpha_{xx}= \alpha_{yy}$), while $M_z$ or $C_{nz}$ forbids the out-of-plane component $\delta s_z$. Overall, the EEE alone operates in 58 (23) MLPGs for the $\mathcal{T}$-even ($\mathcal{T}$-odd) response, and it coexists with the conventional EE in 72 magnetic and 20 nonmagnetic MLPGs \cite{supp_mat}.

%However, the $\mathcal{T}$-odd response (see Table~\ref{tab:odd}) elucidates the intrinsic contribution originating from the band geometry, which exists only in magnetic materials. %Intriguingly, we notice that the conventional EE are not allowed in the 2DMPGs of $-3m'$, $-6m'2'$, $4/mm'm'$, $4'/mm'm$, $6/mm'm'$, $m'm'm(M_z)$, $m'm'm(M_y)$, $m'm'm(M_z)$, $2'/m'(C_{2z}\mathcal{T})$, $2'/m'(C_{2x}\mathcal{T})$ and $2'/m'(C_{2y}\mathcal{T})$, which thus do not have type-II OSP but   support type-I OSP. 
{\color{blue}\textit{Rotating spin via $\mathcal{E}_z$.}}
The EEE plays two distinct roles. When the linear EE is symmetry-forbidden, the EEE alone generates the spin polarization $\delta\bm{s}=\alpha\,\mathcal{E}_z\,\bm{E}$. The gate field controls its magnitude and sign, while its direction is fixed by the EEE tensor and the in-plane field. When the conventional EE is already present, the EEE adds a new, gate-tunable contribution, so that the gate field also rotates the total spin polarization.

When the conventional EE and the EEE coexist, the total spin polarization induced by the crossed in-plane and out-of-plane fields is
\begin{equation}
\delta\bm{s}=\chi\bm{E}+\alpha\,\mathcal{E}_z\,\bm{E}=\big(\chi+\alpha\,\mathcal{E}_z\big)\bm{E},
\end{equation}
where $\chi$ and $\alpha$ are the conventional EE and EEE tensors. At $\mathcal{E}_z=0$, $\delta\bm{s}=\chi\bm{E}$ lies in the plane spanned by the columns of $\chi$. Sweeping the in-plane field rotates $\delta\bm{s}$ within this plane. A finite gate field adds the term $\alpha\,\mathcal{E}_z\,\bm{E}$, which generically tilts $\delta\bm{s}$ out of the plane. For a fixed direction of $\bm{E}$, varying $\mathcal{E}_z$ sweeps $\delta\bm{s}$ along a line in spin space. The direction of $\bm{E}$ and the strength of $\mathcal{E}_z$ therefore provide two independent knobs for the spin polarization.

% The achievable spin polarization is determined by a $4\times3$ matrix assembled from the elements of $\chi$ and $\alpha$,
% \begin{equation}
% M=\begin{pmatrix}
% \chi_{xx} & \chi_{yx} & \chi_{zx}\\
% \chi_{xy} & \chi_{yy} & \chi_{zy}\\
% \alpha_{xx} & \alpha_{yx} & \alpha_{zx}\\
% \alpha_{xy} & \alpha_{yy} & \alpha_{zy}
% \end{pmatrix},
% \end{equation}
% whose four rows correspond to the four channels $\{E_x,\,E_y,\,\mathcal{E}_zE_x,\,\mathcal{E}_zE_y\}$ and whose three columns to the spin components $\{s_x,\,s_y,\,s_z\}$. When $\mathrm{rank}(M)=3$, the four channels span the a large portion of 3D spin space, and the two knobs rotate the spin polarization out of any single plane toward the spin sphere. However, the four channels are not independent. The two EEE channels reduce to the two EE channels multiplied by the common factor $\mathcal{E}_z$. The rotation therefore sweeps almost the entire spin sphere, leaving only very special directions inaccessible. When $\mathrm{rank}(M)=2$, the spin polarization is confined to a single plane, generally not the $x$-$y$ plane but the plane spanned by the columns of $\chi$ and $\alpha$. In this case, only a planar rotation is possible.

To determine the achievable range of spin polarization under the two knobs, we construct a composite response matrix $M$ that combines $\chi$ and $\alpha$. Since $\delta\bm s=\chi\bm E+\alpha\,\mathcal{E}_z\bm E$, the response is fully described by $\delta\bm s=M\bm c$ with $M=(\chi,\ \alpha)$ acting on the four driving channels $\bm c=(E_x,E_y,\mathcal{E}_zE_x,\mathcal{E}_zE_y)^{\mathrm T}$. Explicitly,
\begin{equation}
M=\begin{pmatrix}
\chi_{xx} & \chi_{xy} & \alpha_{xx} & \alpha_{xy}\\
\chi_{yx} & \chi_{yy} & \alpha_{yx} & \alpha_{yy}\\
\chi_{zx} & \chi_{zy} & \alpha_{zx} & \alpha_{zy}
\end{pmatrix}.
\end{equation}
When $\mathrm{rank}(M)=3$, the response spans all three spin directions, and the two knobs rotate the spin polarization out of any single plane. The four driving amplitudes in $\bm c$ are not independently tunable, since the two EEE amplitudes, $\mathcal E_zE_x$ and $\mathcal E_zE_y$, share the same gate field $\mathcal E_z$. The rotation therefore covers a large portion of the spin sphere, with some directions remaining inaccessible. When $\mathrm{rank}(M)=2$, the spin polarization is confined to a single plane, generally not the $x$-$y$ plane but the plane spanned by the columns of $M$. Thus, only a planar rotation of spin polarization is possible.

% {\color{red}In CoSe, $\chi$ and $\alpha$ are both in-plane, so that the gate $\mathcal{E}_z$ rotates the spin within a plane (a rank-2 realization).} By contrast, in CoS $M$ attains rank three, and the spin polarization is rotated over a large portion of the spin sphere---realized for the first time through the Edelstein effect.

\begin{table}[t]
    \centering
    \caption{Constraints on the $\mathcal{T}$-even EEE from MLPG symmetries. ``$\checkmark$'' (``$\times$'') means the component is allowed (forbidden) under the corresponding symmetry operators. Symmetry operators $R\mathcal{T}$ and $R$ impose the same constraints.}\label{tab:even}
    \begin{tabular*}{\linewidth}{@{\extracolsep{\fill}} ccccccccccc @{}}
    
    \hline\hline
    $\alpha^{\text{even}}$ & $\mathcal{P}$ & $C_{2z}$ & $C_{3z/4z/6z}$ & $C_{2x}$ & $C_{2y}$ & $M_z$ & $S_{3z/4z/6z}$ & $M_x$ & $M_y$  \\
    \hline
    $\alpha_{xx}$ & $\checkmark$ & $\checkmark$ & $\checkmark$ & $\times$ & $\times$ & $\checkmark$ & $\checkmark$ & $\times$ & $\times$ \\
    $\alpha_{yy}$ & $\checkmark$ & $\checkmark$ & $\alpha_{xx}$ & $\times$ & $\times$ & $\checkmark$ & $\alpha_{xx}$ & $\times$ & $\times$ \\
    $\alpha_{xy}$ & $\checkmark$ & $\checkmark$ & $\checkmark$ & $\checkmark$ & $\checkmark$ & $\checkmark$ & $\checkmark$ & $\checkmark$ & $\checkmark$ \\
    $\alpha_{yx}$ & $\checkmark$ & $\checkmark$ & $-\alpha_{xy}$ & $\checkmark$ & $\checkmark$ & $\checkmark$ & $-\alpha_{xy}$ & $\checkmark$ & $\checkmark$ \\
    $\alpha_{zx}$ & $\checkmark$ & $\times$ & $\times$ & $\checkmark$ & $\times$ & $\times$ & $\times$ & $\checkmark$ & $\times$ \\
    $\alpha_{zy}$ & $\checkmark$ & $\times$ & $\times$ & $\times$ & $\checkmark$ & $\times$ & $\times$ & $\times$ & $\checkmark$ \\
    \hline\hline
    \end{tabular*}
\end{table}

\begin{table*}[t]
    \centering
    \caption{Constraints on the $\mathcal{T}$-odd EEE from MLPG symmetries. ``$\checkmark$'' (``$\times$'') means the component is allowed (forbidden) under the corresponding symmetry operators. Symmetry operators $\mathcal{T}$, $\mathcal{PT}$, $C_{3z/6z}\mathcal{T}$, and $S_{3z/6z}\mathcal{T}$ forbid all the elements here; they are not listed.}\label{tab:odd}
    \begin{tabular*}{\linewidth}{@{\extracolsep{\fill}} ccccccccccccccccccccc @{}}
    \hline\hline
    $\alpha^{\text{odd}}$ & $\mathcal{P}$ & $C_{2z}$ & $C_{3z/4z/6z}$ & $C_{2x}$ & $C_{2y}$ & $M_z$ & $S_{3z/4z/6z}$ & $M_x$ & $M_y$ & $C_{2z}\mathcal{T}$ & $C_{4z}\mathcal{T}$ & $C_{2x}\mathcal{T}$ & $C_{2y}\mathcal{T}$ & $M_z\mathcal{T}$ & $S_{4z}\mathcal{T}$ & $M_x\mathcal{T}$ & $M_y\mathcal{T}$ \\
    \hline
    $\alpha_{xx}$ & $\checkmark$ & $\checkmark$ & $\checkmark$ & $\times$ & $\times$ & $\checkmark$ & $\checkmark$ & $\times$ & $\times$ & $\times$ & $\checkmark$ & $\checkmark$ & $\checkmark$ & $\times$ & $\checkmark$ & $\checkmark$ & $\checkmark$ \\
    $\alpha_{yy}$ & $\checkmark$ & $\checkmark$ & $\alpha_{xx}$ & $\times$ & $\times$ & $\checkmark$ & $\alpha_{xx}$ &  $\times$ & $\times$ & $\times$ & $-\alpha_{xx}$ & $\checkmark$ & $\checkmark$ & $\times$ & $-\alpha_{xx}$ & $\checkmark$ & $\checkmark$ \\
    $\alpha_{xy}$ & $\checkmark$ & $\checkmark$ & $\checkmark$ & $\checkmark$ & $\checkmark$ & $\checkmark$ & $\checkmark$ & $\checkmark$ & $\checkmark$ & $\times$ & $\checkmark$ & $\times$ & $\times$ & $\times$ & $\checkmark$ & $\times$ & $\times$ \\
    $\alpha_{yx}$ & $\checkmark$ & $\checkmark$ & $-\alpha_{xy}$ & $\checkmark$ & $\checkmark$ & $\checkmark$ & $-\alpha_{xy}$ & $\checkmark$ & $\checkmark$ & $\times$ & $\alpha_{xy}$ & $\times$ & $\times$ & $\times$ & $\alpha_{xy}$ & $\times$ & $\times$ \\
    $\alpha_{zx}$ & $\checkmark$ & $\times$ & $\times$ & $\checkmark$ & $\times$ & $\times$ & $\times$ & $\checkmark$ & $\times$ & $\checkmark$ & $\times$ & $\times$ & $\checkmark$ & $\checkmark$ & $\times$ & $\times$ & $\checkmark$ \\
    $\alpha_{zy}$ & $\checkmark$ & $\times$ & $\times$ & $\times$ & $\checkmark$ & $\times$ & $\times$ & $\times$ & $\checkmark$ & $\checkmark$ & $\times$ & $\checkmark$ & $\times$ & $\checkmark$ & $\times$ & $\checkmark$ & $\times$ \\
    \hline\hline
    \end{tabular*}
\end{table*}

{\color{blue}\textit{Analytical expression of EEE coefficient.}}
The analytical expression of the conventional EE is well established \cite{manchon2019current,awschalom2009spintronics,aronov1989nuclear,edelstein1990spin,fang2011spin,ivchenko1978new,kato2004current,zelezny2014relativistic}, but that of the EEE has not been derived. Such an expression is needed both to understand the microscopic origin of the EEE and to identify materials with a pronounced EEE. We derive it here. To this end, we consider a 2D crystal under the gate field $\mathcal{E}_z$, which enters through the perturbed Hamiltonian
\begin{equation}\label{eq1}
\hat{H} = \hat{H}_0 + g_E \mathcal{E}_z \hat{z}, 
\end{equation}
where $g_E$ is the effective coupling coefficient and $\hat{z}$ denotes the out-of-plane position operator. 
With the perturbed Hamiltonian (\ref{eq1}), the EEE coefficients can be obtained as  (we set  $e = \hbar = 1$) \cite{xiao2023time,dong2020berry,garate2009influence}
\begin{eqnarray}
\alpha^{\text{odd}}_{ab}=
\sum_n\int[d\bm k]\,
\Big[
f^{(1)}(\varepsilon_{n,k})P_{nn}\Upsilon^n_{ab}
+f(\varepsilon_{n,k})\Lambda^{n}_{ab} 
\Big], 
\label{eq:EEE-odd}
\end{eqnarray}  
and 
\begin{eqnarray}
  \alpha^{\text{even}}_{ab}
&=&
-\tau\sum_n\int[d\bm k]\,
\Big[
f^{(2)}(\varepsilon_{n,k})P_{nn}v_b^ns_a^n  \notag \\ 
&+&f^{(1)}(\varepsilon_{n,k})(\partial_{k_b}P_{nn})s_a^n
+f^{(1)}(\varepsilon_{n,k})v_b^n\xi^n_{a}
\Big],
\label{eq:EEE-even}
\end{eqnarray}
where $n$ and $\bm{k}$ are the band index and the wave-vector, respectively. $[d\bm{k}]$ is the short-band notation for $d\bm{k}/(2\pi)^2$, and $f^{(m)}$ denotes the $m$-th derivative of the Fermi-Dirac distribution $f$ with respect to energy. Here, $P_{nm}=g_E\langle u_n|\hat z|u_m\rangle$, $s_a^{nm}=\langle u_n|\hat s_a|u_m\rangle$, and $v_b^{nm}=\langle u_n|\hat v_b|u_m\rangle$ are the layer-polarization, spin, and velocity matrix elements, respectively; $|u_n(\bm k)\rangle$ is the Bloch state of the unperturbed Hamiltonian $\hat H_0$. Hereafter, we set $s_a^n\equiv s_a^{nn}$ and $v_b^n\equiv v_b^{nn}=\partial_{k_b}\varepsilon_n$.

The EEE is therefore governed by three band-geometric quantities: the gate-induced spin polarizability $\xi_a^n=2\operatorname{Re}\sum_{m\ne n}\frac{s_a^{nm}P_{mn}}{\varepsilon_n-\varepsilon_m}$, the anomalous spin polarizability (ASP) $\Upsilon_{ab}^n=2\operatorname{Im}\sum_{m\ne n}\frac{s_a^{nm}v_b^{mn}}{(\varepsilon_n-\varepsilon_m)^2}$, and the gate-induced polarizability of the ASP \cite{xiao2023time},
\begin{eqnarray}
\Lambda^{n}_{ab}
&=&
2\operatorname{Im}\sum_{m\ne n}
\left[
-2\frac{P_{nn}-P_{mm}}
{(\varepsilon_n-\varepsilon_m)^3}
s_a^{nm}v_b^{mn}
\right. \nonumber\\
&&
+\sum_{l\ne m}
\frac{
P_{lm}^{*}v_b^{ln}s_a^{nm}
+P_{lm}s_a^{nl}v_b^{mn}
}
{(\varepsilon_n-\varepsilon_m)^2
(\varepsilon_m-\varepsilon_l)}
\nonumber\\
&&
\left.
+\sum_{l\ne n}
\frac{
P_{ln}v_b^{ml}s_a^{nm}
+P_{ln}^{*}s_a^{lm}v_b^{mn}
}
{(\varepsilon_n-\varepsilon_m)^2
(\varepsilon_n-\varepsilon_l)}
\right].
\label{eq:lambda-main}
\end{eqnarray}

Several significant features emerge from these expressions. First, the $\mathcal{T}$-odd response coefficient $\alpha_{ab}^{\text{odd}}$ is dominated by two band-geometric quantities, $P_{nn}\Upsilon_{ab}^n$ and $\Lambda_{ab}^n$, which may be viewed as the ASP polarization and the ASP polarizability with respect to $\mathcal{E}_z$, respectively. By contrast, the three terms in $\alpha_{ab}^{\text{even}}$ describe gate-induced corrections to the carrier occupation, the band velocity, and the spin texture of the conventional drift-type Edelstein response. Second, $\alpha_{ab}^{\text{odd}}$ is intrinsic and fully determined by the band structure. It contains a Fermi-surface contribution (the first term) and a Fermi-sea contribution from all occupied states (the second term), so it applies to insulators and metals alike. In contrast, $\alpha_{ab}^{\text{even}}$ is extrinsic, and each of its terms is tied to the Fermi surface through $f^{(1)}$ or $f^{(2)}$. It is therefore confined to metals and scales with the scattering time $\tau$. Third, both $\Lambda_{ab}^n$ and $\xi_a^n$, which enter $\alpha_{ab}^{\text{odd}}$ and $\alpha_{ab}^{\text{even}}$, respectively, involve interband matrix elements $P_{nm}$ ($n \neq m$). A finite EEE therefore does not require a significant layer polarization $P_{nn}$ near the Fermi energy. Forth, the small interband energy denominators in $\Upsilon_{ab}^n, \Lambda_{ab}^n$, and $\xi_a^n$ indicates a strong enhancement in the response around small-gap region or topological nodal-points. %This mechanism accounts for the pronounced EEE peaks observed in the material examples discussed below.

More importantly, the EEE is not obtained by simple dimensional reduction from the nonlinear EE formulated for 3D systems. Although both responses are quadratic in the applied field \cite{xiao2022intrinsic,xiao2023time}, they differ in two essential ways. First, the 3D nonlinear EE is quadratic in a single driving field, whereas the EEE is controlled by two independent fields, the gate field $\mathcal{E}_z$ and the in-plane field $\bm E$. The magnitude and sign of the response can therefore be tuned more flexibly, which enriches the control of the spin polarization. Second, in the 3D nonlinear EE both fields drive transport, whereas in the EEE the out-of-plane field couples to the layer polarization and acts as a gate rather than a driving force. This distinction has a direct signature in the coefficients. The first term of $\alpha_{ab}^{\mathrm{odd}}$ and the first two terms of $\alpha_{ab}^{\mathrm{even}}$ originate from the gate-induced energy shift and the resulting change in carrier occupation. These terms are absent in the 3D nonlinear EE. They are unique to the confined 2D electron system and reflect its response to an out-of-plane electric field.

{\color{blue}\textit{EEE in the centrosymmetric ferromagnet MnSe$_2$.}} To verify the existence of the EEE, we first examine a material in which the conventional EE is symmetry-forbidden but the EEE operates. A suitable candidate is 2D MnSe$_2$, a ferromagnetic metal with a Curie temperature of about 264 K \cite{o2018room,vanherck20202d}. In its pristine state, MnSe$_2$ belongs to $\bar{3}m'1$, where the inversion symmetry $\mathcal{P}$ forbids any linear current-induced spin polarization. Fortunately, the gate field $\mathcal{E}_z$ breaks the operations that flip $z$, such as $\mathcal{P}$ and $S_{3z}$. Then, the MLPG is lowered to $3m'1$, which allows the appearance of the EEE. Our symmetry analysis (Tables~\ref{tab:even} and \ref{tab:odd}) yields four nonzero EEE tensor elements, which reduce to two independent ones, $\alpha_{xy}=-\alpha_{yx}$ for the $\mathcal{T}$-even component and $\alpha_{xx}=\alpha_{yy}$ for the $\mathcal{T}$-odd component. For a driving field along $x$, the $\mathcal{T}$-odd and $\mathcal{T}$-even components point along $\hat{\bm{x}}$ and $\hat{\bm{y}}$, respectively, and are therefore mutually orthogonal,
\begin{equation}
\delta\boldsymbol{s}^{\rm odd}=\alpha_{xx}\mathcal{E}_zE_x\,\hat{\bm{x}},\qquad
\delta\boldsymbol{s}^{\rm even}=\alpha_{yx}\mathcal{E}_zE_x\,\hat{\bm{y}}.
\end{equation}
This realizes the first role of the EEE, in which the spin polarization is rotated by the EEE alone.

% \begin{figure}[t]
%         \centering
%         \includegraphics[width=0.5\textwidth]{MnSe2.pdf}
%         \caption{{EEE in the centrosymmetric ferromagnet MnSe$_2$. (a) Top view and (b) side view of 2D MnSe$_2$. The arrows in (b) represent the magnetic moment from Mn atoms.  (c) Band structure of 2D MnSe$_2$ along the high-symmetry paths as SOC is considered. (d) EEE coefficients for the $\mathcal{T}$-even component $\alpha_{xy}$ and the $\mathcal{T}$-odd component $\alpha_{xx}$ as a function of $\mu$. Here g$_E$=0.1 and T = $20$ K.}}
%         \label{MnSe}
%     \end{figure}

\begin{figure}[t]
			\centering
			\includegraphics[width=0.5\textwidth]{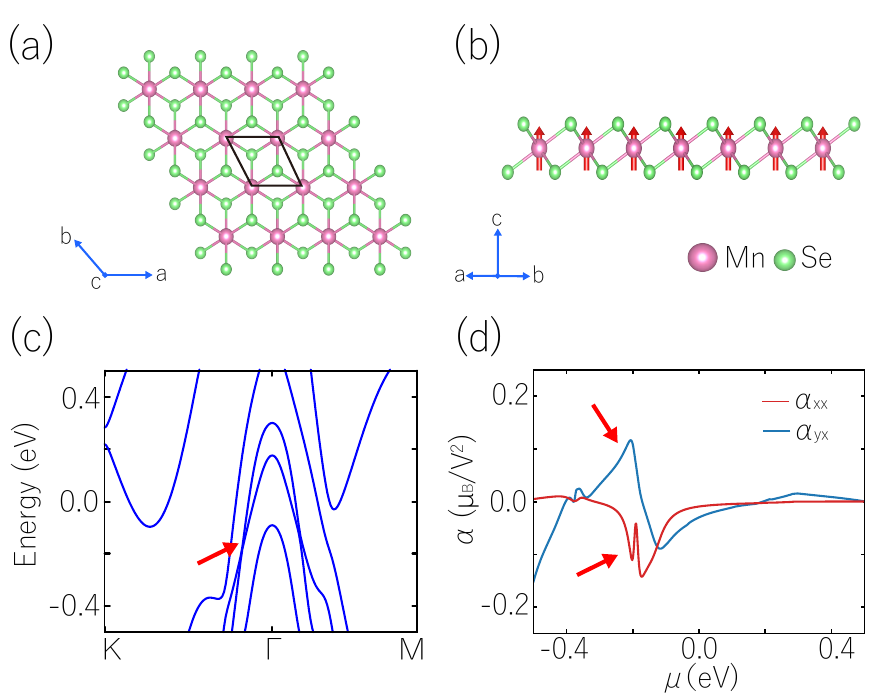}
			\caption{{EEE in the 2D ferromagnetic material MnSe$_2$. (a) Top view and (b) side view of 2D MnSe$_2$. The arrows in (b) represent the magnetic moment from Mn atoms.  (c) Band structure of 2D MnSe$_2$ along the high-symmetry paths as SOC is considered. (d) EEE coefficients for the $\mathcal{T}$-even component $\alpha_{xy}$ and the $\mathcal{T}$-odd component $\alpha_{xx}$ as a function of $\mu$. Here g$_E$=0.1 and T = $20$ K.}}
			\label{MnSe}
            \vspace{-1.5ex}
		\end{figure}

The band structure of 2D MnSe$_2$ with spin-orbit coupling (SOC) included is displayed in Fig.~\ref{MnSe}(c). The system is metallic, with small gaps emerging below the Fermi level. Using Eqs.~(\ref{eq:EEE-odd}) and (\ref{eq:EEE-even}), we calculate the EEE coefficients as a function of the chemical potential $\mu$, taking the relaxation time $\tau=0.05$ ps \cite{ma2019observation,xiao2023time}. Pronounced peaks appear around the small-gap regions (Fig.~\ref{MnSe}(c)-(d)), confirming that the enhancement originates from the small interband energy denominators in the band-geometric quantities. At $\mu=0$ ($-0.2$) eV, we find $\alpha_{yx}=-0.030$ ($0.117$) $\mu_B/V^{2}$ and $\alpha_{xx}=-0.009$ ($-0.099$) $\mu_B/V^{2}$, which give a total spin polarization $\delta s=0.63\times10^{-6}$ ($3.07\times10^{-6}$) $\mu_B/\mathrm{nm}^{2}$ for $\mathcal{E}_z=0.02$ eV/Å and a driving field $E_x=10^{5}$ V/m \cite{chernyshov2009evidence,olejnik2018terahertz,zhou2019fieldlike}. The calculated spin polarization upon hole doping at $\mu$ = $-0.2$ eV are comparable to or even exceed the results reported from the conventional EE \cite{chernyshov2009evidence,kurebayashi2014antidamping}. Moreover, the two mutually orthogonal responses scale differently with $\tau$, and only the $\mathcal{T}$-odd contribution reverses its sign when the magnetization direction is switched. These features thus provide a robust experimental signature that distinguishes the two components.

{\color{blue}\textit{Gate-tunuable 3D spin-polarization rotation in monolayer $SnSe$.}}
We now examine a nonmagnetic material in which the conventional EE and the EEE coexist and can be combined to steer the spin polarization. A suitable example is monolayer SnSe, an experimentally synthesized semiconductor with in-plane ferroelectricity and a large piezoelectric coefficient \cite{shi2015anisotropic,li2013single,fang2016layer}. In its pristine state, SnSe belongs to the MLPG $mm2.1'$, which restricts the conventional EE to a single $\mathcal{T}$-even tensor element $\chi_{zy}$ and forbids the $\mathcal{T}$-odd EE \cite{supp_mat}. The gate field $\mathcal{E}_z$ lowers the MLPG to $m.1'$, which leads to two independent $\mathcal{T}$-even EEE elements, $\alpha_{xy}$ and $\alpha_{yx}$. For an in-plane driving field $\bm{E}=E_0(\cos\theta,\sin\theta)^T$, the two effects induce spin polarizations
\begin{equation}
\begin{split}
\delta\boldsymbol{s}^{\text{EE}} &= \chi_{zy} E_0 \sin\theta\hat{\boldsymbol{z}}, \\
\delta\boldsymbol{s}^{\text{EEE}} &= \mathcal{E}_z E_0 \big(\alpha_{xy}\sin\theta\hat{\boldsymbol{x}} + \alpha_{yx}\cos\theta\hat{\boldsymbol{y}}\big).
\end{split}
\end{equation}
The conventional EE therefore governs only the out-of-plane component $\delta s_z$, while the EEE controls the in-plane components $\delta s_x$ and $\delta s_y$; the two contributions are thus mutually orthogonal. Because the two effects act on orthogonal spin components, the combined response matrix attains $\operatorname{rank}(M)=3$, and the response spans all three spin directions. Notably, the in-plane $\delta s_x$ and the out-of-plane $\delta s_z$ are nevertheless locked by the relation $\delta s_x/\delta s_z=(\alpha_{xy}/\chi_{zy})\mathcal{E}_z$. Under realistic gate-field constraints ($|\mathcal{E}_z|\le\mathcal{E}_{\max}$), the accessible spin polarization therefore covers a broad solid angle on the spin sphere, yet leaving two finite forbidden domains centered at the $\pm\hat{\boldsymbol{x}}$ poles.

\begin{figure}[t]
			\centering
			\includegraphics[width=0.5\textwidth]{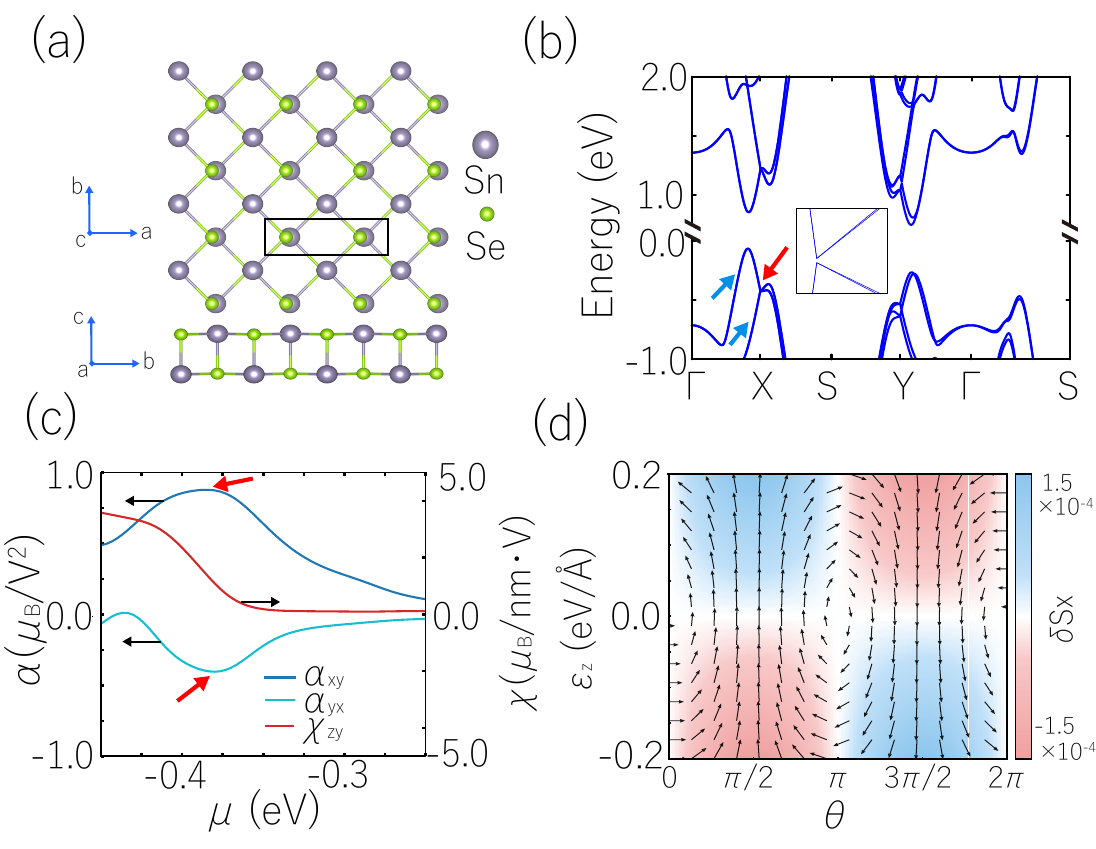}
			\caption{{ Gate-tunable rotation in the 2D SeSn. (a) Top view (upper) and side view (lower) of the monolayer SnSe crystal structure. (b) Electronic band structure with SOC included along the high-symmetry paths. Blue and red  arrows in (b) mark the nodal-lines and the small bang-gap, respectively. (c) The three response coefficients ($\chi_{zy}$, $\alpha_{yx}$ and $\alpha_{xy}$) versus $\mu$. (D) Two-dimensional projections of the non-equilibrium spin accumulation in the $(\theta, \mathcal{E}_z)$ parameter space. The colormap denotes $\delta s_x$ (in a unit of $\mu_B/nm^2$), while the black arrows represent the orthogonal components $(\delta s_y, \delta s_z)$. Here, $g_E = 0.1$ and $T = 100\text{ K}$.}}
			\label{SnSe}
            \vspace{-2ex}
\end{figure}
Fig.~\ref{SnSe}(b) shows the calculated band structure of monolayer SnSe. One observes that it is semiconducting with an indirect band gap, agreeing with the previous results \cite{shi2015anisotropic,fang2016layer}. Interestingly, there exist two doubly degenerate nodal lines running along $\Gamma$-X (blue arrows), and a small band gap of about 0.13 meV opens at the X points (red arrow). Both features are critical for the enhancement of the induced spin polarization. Using Eq.~(\ref{eq:EEE-even}) and its conventional-EE counterpart \cite{manchon2019current}, we evaluate $\alpha_{xy}$, $\alpha_{yx}$, and $\chi_{zy}$ as functions of $\mu$, adopting $\tau=0.05$~ps \cite{ma2019observation,xiao2023time}. As shown in Fig.~\ref{SnSe}(c), the EEE coefficients are enhanced within the energy range of the nodal lines, and they display pronounced peaks around the small band gap at $\mu\approx-0.38$ eV. The resulting total spin accumulation is mapped onto the parameter space $(\theta,\mathcal{E}_z)$ in Fig.~\ref{SnSe}(d). The arrows encode the orthogonal components $\delta s_y$ and $\delta s_z$, and the color map encodes $\delta s_x$, so that each point in the plane represents a complete spin polarization vector. Clearly, varying $\mathcal{E}_z$ and $\theta$ reconfigures the spin orientation continuously. This offers an energy-efficient platform for steering the spin polarization in three-dimensional spin space by purely electrical means.
%{Type-I OSP in the 2D nomagnetic material SnSe. (a) Top view and  side view of 2D SnSe.   (b) Band structure of 2D SnSe along the high-symmetry paths as SOC is considered. (c) EEE and EE coefficients  as a function of $\mu$. Here g$_E$=0.1 and T = $100$ K.(d) Two-dimensional spin texture map, in which the background color represents the X component of spin polarization, and the arrows represent the y-z spin vector distribution.}}

{\color{blue}\textit{Discussion.}} In this work, we have proposed the electric Edelstein effect, wherein crossed out-of-plane and in-plane electric fields generate a gate-tunable spin polarization in 2D systems. A microscopic theory traces the effect to three band-geometric quantities: the ASP polarization, the ASP polarizability, and the gate-induced spin polarizability. All of them are strongly enhanced near small band gaps. Through a symmetry analysis over all MLPGs and first-principles calculations, we have demonstrated the effect in two representative materials. In the centrosymmetric ferromagnet MnSe$_2$, the EEE emerges even though the linear EE is symmetry-forbidden; and in the nonmagnet SnSe,  the combination of the conventional EE and EEE, for the first time, realizes a rotation of the spin polarization over the large portion of spin sphere,  offering an electric knob for multistate spintronics. While our analysis focuses on conventional ferromagnets and nonmagnetic materials, the framework applies equally to antiferromagnet  (e.g., V$_2$Te$_2$O \cite{supp_mat}). Our work therefore establishes the EEE as a symmetry-guided, gate-tunable route to electrical spin control in 2D quantum materials.

%: distinct gate voltages select distinct spin orientations, extending the binary up/down states of a conventional magnet to multilevel memory; the same gate control can also drive spin-orbit torques in magnetic materials. 

%within the detection limits of modern magneto-optical Kerr effect (MOKE) or spin-orbit torque measurements....

\bibliography{ref}

\begin{thebibliography}{48}%
\makeatletter
\providecommand \@ifxundefined [1]{%
 \@ifx{#1\undefined}
}%
\providecommand \@ifnum [1]{%
 \ifnum #1\expandafter \@firstoftwo
 \else \expandafter \@secondoftwo
 \fi
}%
\providecommand \@ifx [1]{%
 \ifx #1\expandafter \@firstoftwo
 \else \expandafter \@secondoftwo
 \fi
}%
\providecommand \natexlab [1]{#1}%
\providecommand \enquote  [1]{``#1''}%
\providecommand \bibnamefont  [1]{#1}%
\providecommand \bibfnamefont [1]{#1}%
\providecommand \citenamefont [1]{#1}%
\providecommand \href@noop [0]{\@secondoftwo}%
\providecommand \href [0]{\begingroup \@sanitize@url \@href}%
\providecommand \@href[1]{\@@startlink{#1}\@@href}%
\providecommand \@@href[1]{\endgroup#1\@@endlink}%
\providecommand \@sanitize@url [0]{\catcode `\\12\catcode `\$12\catcode
  `\&12\catcode `\#12\catcode `\^12\catcode `\_12\catcode `\%12\relax}%
\providecommand \@@startlink[1]{}%
\providecommand \@@endlink[0]{}%
\providecommand \url  [0]{\begingroup\@sanitize@url \@url }%
\providecommand \@url [1]{\endgroup\@href {#1}{\urlprefix }}%
\providecommand \urlprefix  [0]{URL }%
\providecommand \Eprint [0]{\href }%
\providecommand \doibase [0]{http://dx.doi.org/}%
\providecommand \selectlanguage [0]{\@gobble}%
\providecommand \bibinfo  [0]{\@secondoftwo}%
\providecommand \bibfield  [0]{\@secondoftwo}%
\providecommand \translation [1]{[#1]}%
\providecommand \BibitemOpen [0]{}%
\providecommand \bibitemStop [0]{}%
\providecommand \bibitemNoStop [0]{.\EOS\space}%
\providecommand \EOS [0]{\spacefactor3000\relax}%
\providecommand \BibitemShut  [1]{\csname bibitem#1\endcsname}%
\let\auto@bib@innerbib\@empty
%</preamble>
\bibitem [{\citenamefont {Manchon}\ \emph {et~al.}(2019)\citenamefont
  {Manchon}, \citenamefont {{\v{Z}}elezn{\'y}}, \citenamefont {Miron},
  \citenamefont {Jungwirth}, \citenamefont {Sinova}, \citenamefont {Thiaville},
  \citenamefont {Garello},\ and\ \citenamefont
  {Gambardella}}]{manchon2019current}%
  \BibitemOpen
  \bibfield  {author} {\bibinfo {author} {\bibfnamefont {A.}~\bibnamefont
  {Manchon}}, \bibinfo {author} {\bibfnamefont {J.}~\bibnamefont
  {{\v{Z}}elezn{\'y}}}, \bibinfo {author} {\bibfnamefont {I.~M.}\ \bibnamefont
  {Miron}}, \bibinfo {author} {\bibfnamefont {T.}~\bibnamefont {Jungwirth}},
  \bibinfo {author} {\bibfnamefont {J.}~\bibnamefont {Sinova}}, \bibinfo
  {author} {\bibfnamefont {A.}~\bibnamefont {Thiaville}}, \bibinfo {author}
  {\bibfnamefont {K.}~\bibnamefont {Garello}}, \ and\ \bibinfo {author}
  {\bibfnamefont {P.}~\bibnamefont {Gambardella}},\ }\href {\doibase
  10.1103/RevModPhys.91.035004} {\bibfield  {journal} {\bibinfo  {journal}
  {Rev. Mod. Phys.}\ }\textbf {\bibinfo {volume} {91}},\ \bibinfo {pages}
  {035004} (\bibinfo {year} {2019})}\BibitemShut {NoStop}%
\bibitem [{\citenamefont {Awschalom}\ and\ \citenamefont
  {Samarth}(2009)}]{awschalom2009spintronics}%
  \BibitemOpen
  \bibfield  {author} {\bibinfo {author} {\bibfnamefont {D.}~\bibnamefont
  {Awschalom}}\ and\ \bibinfo {author} {\bibfnamefont {N.}~\bibnamefont
  {Samarth}},\ }\href {\doibase 10.1103/Physics.2.50} {\bibfield  {journal}
  {\bibinfo  {journal} {Physics}\ }\textbf {\bibinfo {volume} {2}},\ \bibinfo
  {pages} {50} (\bibinfo {year} {2009})}\BibitemShut {NoStop}%
\bibitem [{\citenamefont {Aronov}\ and\ \citenamefont
  {Lyanda-Geller}(1989)}]{aronov1989nuclear}%
  \BibitemOpen
  \bibfield  {author} {\bibinfo {author} {\bibfnamefont {A.~G.}\ \bibnamefont
  {Aronov}}\ and\ \bibinfo {author} {\bibfnamefont {Y.~B.}\ \bibnamefont
  {Lyanda-Geller}},\ }\href@noop {} {\bibfield  {journal} {\bibinfo  {journal}
  {JETP Lett.}\ }\textbf {\bibinfo {volume} {50}},\ \bibinfo {pages} {431}
  (\bibinfo {year} {1989})}\BibitemShut {NoStop}%
\bibitem [{\citenamefont {Edelstein}(1990)}]{edelstein1990spin}%
  \BibitemOpen
  \bibfield  {author} {\bibinfo {author} {\bibfnamefont {V.~M.}\ \bibnamefont
  {Edelstein}},\ }\href {\doibase 10.1016/0038-1098(90)90963-C} {\bibfield
  {journal} {\bibinfo  {journal} {Solid State Commun.}\ }\textbf {\bibinfo
  {volume} {73}},\ \bibinfo {pages} {233} (\bibinfo {year} {1990})}\BibitemShut
  {NoStop}%
\bibitem [{\citenamefont {Fang}\ \emph {et~al.}(2011)\citenamefont {Fang},
  \citenamefont {Kurebayashi}, \citenamefont {Wunderlich}, \citenamefont
  {V{\'y}born{\'y}}, \citenamefont {Z{\^a}rbo}, \citenamefont {Campion},
  \citenamefont {Casiraghi}, \citenamefont {Gallagher}, \citenamefont
  {Jungwirth},\ and\ \citenamefont {Ferguson}}]{fang2011spin}%
  \BibitemOpen
  \bibfield  {author} {\bibinfo {author} {\bibfnamefont {D.}~\bibnamefont
  {Fang}}, \bibinfo {author} {\bibfnamefont {H.}~\bibnamefont {Kurebayashi}},
  \bibinfo {author} {\bibfnamefont {J.}~\bibnamefont {Wunderlich}}, \bibinfo
  {author} {\bibfnamefont {K.}~\bibnamefont {V{\'y}born{\'y}}}, \bibinfo
  {author} {\bibfnamefont {L.~P.}\ \bibnamefont {Z{\^a}rbo}}, \bibinfo {author}
  {\bibfnamefont {R.~P.}\ \bibnamefont {Campion}}, \bibinfo {author}
  {\bibfnamefont {A.}~\bibnamefont {Casiraghi}}, \bibinfo {author}
  {\bibfnamefont {B.~L.}\ \bibnamefont {Gallagher}}, \bibinfo {author}
  {\bibfnamefont {T.}~\bibnamefont {Jungwirth}}, \ and\ \bibinfo {author}
  {\bibfnamefont {A.~J.}\ \bibnamefont {Ferguson}},\ }\href {\doibase
  10.1038/nnano.2011.68} {\bibfield  {journal} {\bibinfo  {journal} {Nat.
  Nanotechnol.}\ }\textbf {\bibinfo {volume} {6}},\ \bibinfo {pages} {413}
  (\bibinfo {year} {2011})}\BibitemShut {NoStop}%
\bibitem [{\citenamefont {Ivchenko}\ and\ \citenamefont
  {Pikus}(1978)}]{ivchenko1978new}%
  \BibitemOpen
  \bibfield  {author} {\bibinfo {author} {\bibfnamefont {E.~L.}\ \bibnamefont
  {Ivchenko}}\ and\ \bibinfo {author} {\bibfnamefont {G.~E.}\ \bibnamefont
  {Pikus}},\ }\href@noop {} {\bibfield  {journal} {\bibinfo  {journal} {JETP
  Lett.}\ }\textbf {\bibinfo {volume} {27}},\ \bibinfo {pages} {604} (\bibinfo
  {year} {1978})}\BibitemShut {NoStop}%
\bibitem [{\citenamefont {Kato}\ \emph {et~al.}(2004)\citenamefont {Kato},
  \citenamefont {Myers}, \citenamefont {Gossard},\ and\ \citenamefont
  {Awschalom}}]{kato2004current}%
  \BibitemOpen
  \bibfield  {author} {\bibinfo {author} {\bibfnamefont {Y.~K.}\ \bibnamefont
  {Kato}}, \bibinfo {author} {\bibfnamefont {R.~C.}\ \bibnamefont {Myers}},
  \bibinfo {author} {\bibfnamefont {A.~C.}\ \bibnamefont {Gossard}}, \ and\
  \bibinfo {author} {\bibfnamefont {D.~D.}\ \bibnamefont {Awschalom}},\ }\href
  {\doibase 10.1103/PhysRevLett.93.176601} {\bibfield  {journal} {\bibinfo
  {journal} {Phys. Rev. Lett.}\ }\textbf {\bibinfo {volume} {93}},\ \bibinfo
  {pages} {176601} (\bibinfo {year} {2004})}\BibitemShut {NoStop}%
\bibitem [{\citenamefont {{\v{Z}}elezn{\'y}}\ \emph {et~al.}(2014)\citenamefont
  {{\v{Z}}elezn{\'y}}, \citenamefont {Gao}, \citenamefont {V{\'y}born{\'y}},
  \citenamefont {Zemen}, \citenamefont {Ma{\v{s}}ek}, \citenamefont {Manchon},
  \citenamefont {Wunderlich}, \citenamefont {Sinova},\ and\ \citenamefont
  {Jungwirth}}]{zelezny2014relativistic}%
  \BibitemOpen
  \bibfield  {author} {\bibinfo {author} {\bibfnamefont {J.}~\bibnamefont
  {{\v{Z}}elezn{\'y}}}, \bibinfo {author} {\bibfnamefont {H.}~\bibnamefont
  {Gao}}, \bibinfo {author} {\bibfnamefont {K.}~\bibnamefont
  {V{\'y}born{\'y}}}, \bibinfo {author} {\bibfnamefont {J.}~\bibnamefont
  {Zemen}}, \bibinfo {author} {\bibfnamefont {J.}~\bibnamefont {Ma{\v{s}}ek}},
  \bibinfo {author} {\bibfnamefont {A.}~\bibnamefont {Manchon}}, \bibinfo
  {author} {\bibfnamefont {J.}~\bibnamefont {Wunderlich}}, \bibinfo {author}
  {\bibfnamefont {J.}~\bibnamefont {Sinova}}, \ and\ \bibinfo {author}
  {\bibfnamefont {T.}~\bibnamefont {Jungwirth}},\ }\href {\doibase
  10.1103/PhysRevLett.113.157201} {\bibfield  {journal} {\bibinfo  {journal}
  {Phys. Rev. Lett.}\ }\textbf {\bibinfo {volume} {113}},\ \bibinfo {pages}
  {157201} (\bibinfo {year} {2014})}\BibitemShut {NoStop}%
\bibitem [{\citenamefont {Zhou}\ \emph {et~al.}(2026)\citenamefont {Zhou},
  \citenamefont {Zhou}, \citenamefont {Pan}, \citenamefont {Hu}, \citenamefont
  {Liu},\ and\ \citenamefont {Sun}}]{zhou2026layer}%
  \BibitemOpen
  \bibfield  {author} {\bibinfo {author} {\bibfnamefont {B.}~\bibnamefont
  {Zhou}}, \bibinfo {author} {\bibfnamefont {P.}~\bibnamefont {Zhou}}, \bibinfo
  {author} {\bibfnamefont {B.}~\bibnamefont {Pan}}, \bibinfo {author}
  {\bibfnamefont {Y.}~\bibnamefont {Hu}}, \bibinfo {author} {\bibfnamefont
  {S.}~\bibnamefont {Liu}}, \ and\ \bibinfo {author} {\bibfnamefont
  {L.}~\bibnamefont {Sun}},\ }\href {\doibase 10.1103/PhysRevLett.137.076401}
  {\bibfield  {journal} {\bibinfo  {journal} {Phys. Rev. Lett.}\ }\textbf
  {\bibinfo {volume} {137}},\ \bibinfo {pages} {076401} (\bibinfo {year}
  {2026})}\BibitemShut {NoStop}%
\bibitem [{\citenamefont {Miron}\ \emph {et~al.}(2010)\citenamefont {Miron},
  \citenamefont {Gaudin}, \citenamefont {Auffret}, \citenamefont {Rodmacq},
  \citenamefont {Schuhl}, \citenamefont {Pizzini}, \citenamefont {Vogel},\ and\
  \citenamefont {Gambardella}}]{mihai2010current}%
  \BibitemOpen
  \bibfield  {author} {\bibinfo {author} {\bibfnamefont {I.~M.}\ \bibnamefont
  {Miron}}, \bibinfo {author} {\bibfnamefont {G.}~\bibnamefont {Gaudin}},
  \bibinfo {author} {\bibfnamefont {S.}~\bibnamefont {Auffret}}, \bibinfo
  {author} {\bibfnamefont {B.}~\bibnamefont {Rodmacq}}, \bibinfo {author}
  {\bibfnamefont {A.}~\bibnamefont {Schuhl}}, \bibinfo {author} {\bibfnamefont
  {S.}~\bibnamefont {Pizzini}}, \bibinfo {author} {\bibfnamefont
  {J.}~\bibnamefont {Vogel}}, \ and\ \bibinfo {author} {\bibfnamefont
  {P.}~\bibnamefont {Gambardella}},\ }\href {\doibase 10.1038/nmat2613}
  {\bibfield  {journal} {\bibinfo  {journal} {Nat. Mater.}\ }\textbf {\bibinfo
  {volume} {9}},\ \bibinfo {pages} {230} (\bibinfo {year} {2010})}\BibitemShut
  {NoStop}%
\bibitem [{\citenamefont {Miron}\ \emph {et~al.}(2011)\citenamefont {Miron},
  \citenamefont {Garello}, \citenamefont {Gaudin}, \citenamefont {Zermatten},
  \citenamefont {Costache}, \citenamefont {Auffret}, \citenamefont {Bandiera},
  \citenamefont {Rodmacq}, \citenamefont {Schuhl},\ and\ \citenamefont
  {Gambardella}}]{miron2011perpendicular}%
  \BibitemOpen
  \bibfield  {author} {\bibinfo {author} {\bibfnamefont {I.~M.}\ \bibnamefont
  {Miron}}, \bibinfo {author} {\bibfnamefont {K.}~\bibnamefont {Garello}},
  \bibinfo {author} {\bibfnamefont {G.}~\bibnamefont {Gaudin}}, \bibinfo
  {author} {\bibfnamefont {P.-J.}\ \bibnamefont {Zermatten}}, \bibinfo {author}
  {\bibfnamefont {M.~V.}\ \bibnamefont {Costache}}, \bibinfo {author}
  {\bibfnamefont {S.}~\bibnamefont {Auffret}}, \bibinfo {author} {\bibfnamefont
  {S.}~\bibnamefont {Bandiera}}, \bibinfo {author} {\bibfnamefont
  {B.}~\bibnamefont {Rodmacq}}, \bibinfo {author} {\bibfnamefont
  {A.}~\bibnamefont {Schuhl}}, \ and\ \bibinfo {author} {\bibfnamefont
  {P.}~\bibnamefont {Gambardella}},\ }\href {\doibase 10.1038/nature10309}
  {\bibfield  {journal} {\bibinfo  {journal} {Nature}\ }\textbf {\bibinfo
  {volume} {476}},\ \bibinfo {pages} {189} (\bibinfo {year}
  {2011})}\BibitemShut {NoStop}%
\bibitem [{\citenamefont {Liu}\ \emph {et~al.}(2012)\citenamefont {Liu},
  \citenamefont {Pai}, \citenamefont {Li}, \citenamefont {Tseng}, \citenamefont
  {Ralph},\ and\ \citenamefont {Buhrman}}]{liu2012spin}%
  \BibitemOpen
  \bibfield  {author} {\bibinfo {author} {\bibfnamefont {L.}~\bibnamefont
  {Liu}}, \bibinfo {author} {\bibfnamefont {C.-F.}\ \bibnamefont {Pai}},
  \bibinfo {author} {\bibfnamefont {Y.}~\bibnamefont {Li}}, \bibinfo {author}
  {\bibfnamefont {H.~W.}\ \bibnamefont {Tseng}}, \bibinfo {author}
  {\bibfnamefont {D.~C.}\ \bibnamefont {Ralph}}, \ and\ \bibinfo {author}
  {\bibfnamefont {R.~A.}\ \bibnamefont {Buhrman}},\ }\href {\doibase
  10.1126/science.1218197} {\bibfield  {journal} {\bibinfo  {journal}
  {Science}\ }\textbf {\bibinfo {volume} {336}},\ \bibinfo {pages} {555}
  (\bibinfo {year} {2012})}\BibitemShut {NoStop}%
\bibitem [{\citenamefont {Freimuth}\ \emph {et~al.}(2014)\citenamefont
  {Freimuth}, \citenamefont {Bl{\"u}gel},\ and\ \citenamefont
  {Mokrousov}}]{freimuth2014spin}%
  \BibitemOpen
  \bibfield  {author} {\bibinfo {author} {\bibfnamefont {F.}~\bibnamefont
  {Freimuth}}, \bibinfo {author} {\bibfnamefont {S.}~\bibnamefont
  {Bl{\"u}gel}}, \ and\ \bibinfo {author} {\bibfnamefont {Y.}~\bibnamefont
  {Mokrousov}},\ }\href {\doibase 10.1103/PhysRevB.90.174423} {\bibfield
  {journal} {\bibinfo  {journal} {Phys. Rev. B}\ }\textbf {\bibinfo {volume}
  {90}},\ \bibinfo {pages} {174423} (\bibinfo {year} {2014})}\BibitemShut
  {NoStop}%
\bibitem [{\citenamefont {{\v{Z}}elezn{\'y}}\ \emph {et~al.}(2017)\citenamefont
  {{\v{Z}}elezn{\'y}}, \citenamefont {Gao}, \citenamefont {Manchon},
  \citenamefont {Freimuth}, \citenamefont {Mokrousov}, \citenamefont {Zemen},
  \citenamefont {Ma{\v{s}}ek}, \citenamefont {Sinova},\ and\ \citenamefont
  {Jungwirth}}]{vzelezny2017spin}%
  \BibitemOpen
  \bibfield  {author} {\bibinfo {author} {\bibfnamefont {J.}~\bibnamefont
  {{\v{Z}}elezn{\'y}}}, \bibinfo {author} {\bibfnamefont {H.}~\bibnamefont
  {Gao}}, \bibinfo {author} {\bibfnamefont {A.}~\bibnamefont {Manchon}},
  \bibinfo {author} {\bibfnamefont {F.}~\bibnamefont {Freimuth}}, \bibinfo
  {author} {\bibfnamefont {Y.}~\bibnamefont {Mokrousov}}, \bibinfo {author}
  {\bibfnamefont {J.}~\bibnamefont {Zemen}}, \bibinfo {author} {\bibfnamefont
  {J.}~\bibnamefont {Ma{\v{s}}ek}}, \bibinfo {author} {\bibfnamefont
  {J.}~\bibnamefont {Sinova}}, \ and\ \bibinfo {author} {\bibfnamefont
  {T.}~\bibnamefont {Jungwirth}},\ }\href {\doibase 10.1103/PhysRevB.95.014403}
  {\bibfield  {journal} {\bibinfo  {journal} {Phys. Rev. B}\ }\textbf {\bibinfo
  {volume} {95}},\ \bibinfo {pages} {014403} (\bibinfo {year}
  {2017})}\BibitemShut {NoStop}%
\bibitem [{\citenamefont {Culcer}\ and\ \citenamefont
  {Winkler}(2007)}]{culcer2007generation}%
  \BibitemOpen
  \bibfield  {author} {\bibinfo {author} {\bibfnamefont {D.}~\bibnamefont
  {Culcer}}\ and\ \bibinfo {author} {\bibfnamefont {R.}~\bibnamefont
  {Winkler}},\ }\href {\doibase 10.1103/PhysRevLett.99.226601} {\bibfield
  {journal} {\bibinfo  {journal} {Phys. Rev. Lett.}\ }\textbf {\bibinfo
  {volume} {99}},\ \bibinfo {pages} {226601} (\bibinfo {year}
  {2007})}\BibitemShut {NoStop}%
\bibitem [{\citenamefont {Xiao}\ \emph {et~al.}(2022)\citenamefont {Xiao},
  \citenamefont {Liu}, \citenamefont {Wu}, \citenamefont {Wang}, \citenamefont
  {Niu},\ and\ \citenamefont {Yang}}]{xiao2022intrinsic}%
  \BibitemOpen
  \bibfield  {author} {\bibinfo {author} {\bibfnamefont {C.}~\bibnamefont
  {Xiao}}, \bibinfo {author} {\bibfnamefont {H.}~\bibnamefont {Liu}}, \bibinfo
  {author} {\bibfnamefont {W.}~\bibnamefont {Wu}}, \bibinfo {author}
  {\bibfnamefont {H.}~\bibnamefont {Wang}}, \bibinfo {author} {\bibfnamefont
  {Q.}~\bibnamefont {Niu}}, \ and\ \bibinfo {author} {\bibfnamefont {S.~Y.~A.}\
  \bibnamefont {Yang}},\ }\href {\doibase 10.1103/PhysRevLett.129.086602}
  {\bibfield  {journal} {\bibinfo  {journal} {Phys. Rev. Lett.}\ }\textbf
  {\bibinfo {volume} {129}},\ \bibinfo {pages} {086602} (\bibinfo {year}
  {2022})}\BibitemShut {NoStop}%
\bibitem [{\citenamefont {Xiao}\ \emph {et~al.}(2023)\citenamefont {Xiao},
  \citenamefont {Wu}, \citenamefont {Wang}, \citenamefont {Huang},
  \citenamefont {Feng}, \citenamefont {Liu}, \citenamefont {Guo}, \citenamefont
  {Niu},\ and\ \citenamefont {Yang}}]{xiao2023time}%
  \BibitemOpen
  \bibfield  {author} {\bibinfo {author} {\bibfnamefont {C.}~\bibnamefont
  {Xiao}}, \bibinfo {author} {\bibfnamefont {W.}~\bibnamefont {Wu}}, \bibinfo
  {author} {\bibfnamefont {H.}~\bibnamefont {Wang}}, \bibinfo {author}
  {\bibfnamefont {Y.-X.}\ \bibnamefont {Huang}}, \bibinfo {author}
  {\bibfnamefont {X.}~\bibnamefont {Feng}}, \bibinfo {author} {\bibfnamefont
  {H.}~\bibnamefont {Liu}}, \bibinfo {author} {\bibfnamefont {G.-Y.}\
  \bibnamefont {Guo}}, \bibinfo {author} {\bibfnamefont {Q.}~\bibnamefont
  {Niu}}, \ and\ \bibinfo {author} {\bibfnamefont {S.~A.}\ \bibnamefont
  {Yang}},\ }\href {\doibase 10.1103/PhysRevLett.130.166302} {\bibfield
  {journal} {\bibinfo  {journal} {Phys. Rev. Lett.}\ }\textbf {\bibinfo
  {volume} {130}},\ \bibinfo {pages} {166302} (\bibinfo {year}
  {2023})}\BibitemShut {NoStop}%
\bibitem [{\citenamefont {Xu}\ \emph {et~al.}(2021)\citenamefont {Xu},
  \citenamefont {Zhou}, \citenamefont {Wang},\ and\ \citenamefont
  {Li}}]{xu2021light}%
  \BibitemOpen
  \bibfield  {author} {\bibinfo {author} {\bibfnamefont {H.~W.}\ \bibnamefont
  {Xu}}, \bibinfo {author} {\bibfnamefont {J.}~\bibnamefont {Zhou}}, \bibinfo
  {author} {\bibfnamefont {H.}~\bibnamefont {Wang}}, \ and\ \bibinfo {author}
  {\bibfnamefont {J.}~\bibnamefont {Li}},\ }\href {\doibase
  10.1103/PhysRevB.103.205417} {\bibfield  {journal} {\bibinfo  {journal}
  {Phys. Rev. B}\ }\textbf {\bibinfo {volume} {103}},\ \bibinfo {pages}
  {205417} (\bibinfo {year} {2021})}\BibitemShut {NoStop}%
\bibitem [{\citenamefont {Xu}\ and\ \citenamefont {Li}(2025)}]{xu2025linear}%
  \BibitemOpen
  \bibfield  {author} {\bibinfo {author} {\bibfnamefont {H.~W.}\ \bibnamefont
  {Xu}}\ and\ \bibinfo {author} {\bibfnamefont {J.}~\bibnamefont {Li}},\ }\href
  {\doibase 10.1016/j.mtquant.2024.100022} {\bibfield  {journal} {\bibinfo
  {journal} {Mater. Today Quantum}\ }\textbf {\bibinfo {volume} {5}},\ \bibinfo
  {pages} {100022} (\bibinfo {year} {2025})}\BibitemShut {NoStop}%
\bibitem [{\citenamefont {Sarkar}\ \emph {et~al.}(2025)\citenamefont {Sarkar},
  \citenamefont {Das}, \citenamefont {Mandal},\ and\ \citenamefont
  {Agarwal}}]{sarkar2025light}%
  \BibitemOpen
  \bibfield  {author} {\bibinfo {author} {\bibfnamefont {S.}~\bibnamefont
  {Sarkar}}, \bibinfo {author} {\bibfnamefont {S.}~\bibnamefont {Das}},
  \bibinfo {author} {\bibfnamefont {D.}~\bibnamefont {Mandal}}, \ and\ \bibinfo
  {author} {\bibfnamefont {A.}~\bibnamefont {Agarwal}},\ }\href {\doibase
  10.1088/1367-2630/ad8e0a} {\bibfield  {journal} {\bibinfo  {journal} {New J.
  Phys.}\ }\textbf {\bibinfo {volume} {27}},\ \bibinfo {pages} {113503}
  (\bibinfo {year} {2025})}\BibitemShut {NoStop}%
\bibitem [{\citenamefont {Feng}\ \emph {et~al.}(2025)\citenamefont {Feng},
  \citenamefont {Wu}, \citenamefont {Wang}, \citenamefont {Gao}, \citenamefont
  {Ang}, \citenamefont {Zhao}, \citenamefont {Xiao},\ and\ \citenamefont
  {Yang}}]{feng2025quantum}%
  \BibitemOpen
  \bibfield  {author} {\bibinfo {author} {\bibfnamefont {X.}~\bibnamefont
  {Feng}}, \bibinfo {author} {\bibfnamefont {W.}~\bibnamefont {Wu}}, \bibinfo
  {author} {\bibfnamefont {H.}~\bibnamefont {Wang}}, \bibinfo {author}
  {\bibfnamefont {W.}~\bibnamefont {Gao}}, \bibinfo {author} {\bibfnamefont
  {L.~K.}\ \bibnamefont {Ang}}, \bibinfo {author} {\bibfnamefont {Y.~X.}\
  \bibnamefont {Zhao}}, \bibinfo {author} {\bibfnamefont {C.}~\bibnamefont
  {Xiao}}, \ and\ \bibinfo {author} {\bibfnamefont {S.~Y.~A.}\ \bibnamefont
  {Yang}},\ }\href {\doibase 10.1016/j.mtquant.2024.100040} {\bibfield
  {journal} {\bibinfo  {journal} {Mater. Today Quantum}\ }\textbf {\bibinfo
  {volume} {6}},\ \bibinfo {pages} {100040} (\bibinfo {year}
  {2025})}\BibitemShut {NoStop}%
\bibitem [{\citenamefont {Cao}\ \emph {et~al.}(2026)\citenamefont {Cao},
  \citenamefont {Wu}, \citenamefont {Liu}, \citenamefont {Lai}, \citenamefont
  {Xiao}, \citenamefont {Niu}, \citenamefont {Xie},\ and\ \citenamefont
  {Yang}}]{cao2026electrical}%
  \BibitemOpen
  \bibfield  {author} {\bibinfo {author} {\bibfnamefont {J.}~\bibnamefont
  {Cao}}, \bibinfo {author} {\bibfnamefont {W.}~\bibnamefont {Wu}}, \bibinfo
  {author} {\bibfnamefont {H.}~\bibnamefont {Liu}}, \bibinfo {author}
  {\bibfnamefont {S.}~\bibnamefont {Lai}}, \bibinfo {author} {\bibfnamefont
  {C.}~\bibnamefont {Xiao}}, \bibinfo {author} {\bibfnamefont {Q.}~\bibnamefont
  {Niu}}, \bibinfo {author} {\bibfnamefont {X.~C.}\ \bibnamefont {Xie}}, \ and\
  \bibinfo {author} {\bibfnamefont {S.~A.}\ \bibnamefont {Yang}},\ }\href
  {\doibase 10.1088/1361-6633/aea365} {\bibfield  {journal} {\bibinfo
  {journal} {Rep. Prog. Phys.}\ ,\ \bibinfo {pages} {aea365}} (\bibinfo {year}
  {2026})}\BibitemShut {NoStop}%
\bibitem [{\citenamefont {Xiang}\ \emph {et~al.}(2025)\citenamefont {Xiang},
  \citenamefont {Jin},\ and\ \citenamefont {Wang}}]{xiang2025spin}%
  \BibitemOpen
  \bibfield  {author} {\bibinfo {author} {\bibfnamefont {L.}~\bibnamefont
  {Xiang}}, \bibinfo {author} {\bibfnamefont {H.}~\bibnamefont {Jin}}, \ and\
  \bibinfo {author} {\bibfnamefont {J.}~\bibnamefont {Wang}},\ }\href {\doibase
  10.1103/PhysRevLett.135.146303} {\bibfield  {journal} {\bibinfo  {journal}
  {Phys. Rev. Lett.}\ }\textbf {\bibinfo {volume} {135}},\ \bibinfo {pages}
  {146303} (\bibinfo {year} {2025})}\BibitemShut {NoStop}%
\bibitem [{\citenamefont {Jia}\ \emph {et~al.}(2026)\citenamefont {Jia},
  \citenamefont {Xiang}, \citenamefont {Qiao},\ and\ \citenamefont
  {Wang}}]{jia2026nonlinear}%
  \BibitemOpen
  \bibfield  {author} {\bibinfo {author} {\bibfnamefont {J.}~\bibnamefont
  {Jia}}, \bibinfo {author} {\bibfnamefont {L.}~\bibnamefont {Xiang}}, \bibinfo
  {author} {\bibfnamefont {Z.}~\bibnamefont {Qiao}}, \ and\ \bibinfo {author}
  {\bibfnamefont {J.}~\bibnamefont {Wang}},\ }\href {\doibase
  10.1103/PhysRevLett.136.226302} {\bibfield  {journal} {\bibinfo  {journal}
  {Phys. Rev. Lett.}\ }\textbf {\bibinfo {volume} {136}},\ \bibinfo {pages}
  {226302} (\bibinfo {year} {2026})}\BibitemShut {NoStop}%
\bibitem [{\citenamefont {Cui}\ \emph {et~al.}(2025)\citenamefont {Cui},
  \citenamefont {Zhang}, \citenamefont {Qiu}, \citenamefont {Han},
  \citenamefont {Yu},\ and\ \citenamefont {Yao}}]{cui2025electric}%
  \BibitemOpen
  \bibfield  {author} {\bibinfo {author} {\bibfnamefont {C.}~\bibnamefont
  {Cui}}, \bibinfo {author} {\bibfnamefont {R.-W.}\ \bibnamefont {Zhang}},
  \bibinfo {author} {\bibfnamefont {Y.}~\bibnamefont {Qiu}}, \bibinfo {author}
  {\bibfnamefont {Y.}~\bibnamefont {Han}}, \bibinfo {author} {\bibfnamefont
  {Z.-M.}\ \bibnamefont {Yu}}, \ and\ \bibinfo {author} {\bibfnamefont
  {Y.}~\bibnamefont {Yao}},\ }\href {\doibase 10.1103/PhysRevLett.135.116301}
  {\bibfield  {journal} {\bibinfo  {journal} {Phys. Rev. Lett.}\ }\textbf
  {\bibinfo {volume} {135}},\ \bibinfo {pages} {116301} (\bibinfo {year}
  {2025})}\BibitemShut {NoStop}%
\bibitem [{\citenamefont {Sahoo}\ \emph {et~al.}(2005)\citenamefont {Sahoo},
  \citenamefont {Kontos}, \citenamefont {Furer}, \citenamefont {Hoffmann},
  \citenamefont {Gr{\"a}ber}, \citenamefont {Cottet},\ and\ \citenamefont
  {Sch{\"o}nenberger}}]{sahoo2005electric}%
  \BibitemOpen
  \bibfield  {author} {\bibinfo {author} {\bibfnamefont {S.}~\bibnamefont
  {Sahoo}}, \bibinfo {author} {\bibfnamefont {T.}~\bibnamefont {Kontos}},
  \bibinfo {author} {\bibfnamefont {J.}~\bibnamefont {Furer}}, \bibinfo
  {author} {\bibfnamefont {C.}~\bibnamefont {Hoffmann}}, \bibinfo {author}
  {\bibfnamefont {M.}~\bibnamefont {Gr{\"a}ber}}, \bibinfo {author}
  {\bibfnamefont {A.}~\bibnamefont {Cottet}}, \ and\ \bibinfo {author}
  {\bibfnamefont {C.}~\bibnamefont {Sch{\"o}nenberger}},\ }\href {\doibase
  10.1038/nphys120} {\bibfield  {journal} {\bibinfo  {journal} {Nat. Phys.}\
  }\textbf {\bibinfo {volume} {1}},\ \bibinfo {pages} {99} (\bibinfo {year}
  {2005})}\BibitemShut {NoStop}%
\bibitem [{\citenamefont {Rovillain}\ \emph {et~al.}(2010)\citenamefont
  {Rovillain}, \citenamefont {De~Sousa}, \citenamefont {Gallais}, \citenamefont
  {Sacuto}, \citenamefont {M{\'e}asson}, \citenamefont {Colson}, \citenamefont
  {Forget}, \citenamefont {Bibes}, \citenamefont {Barth{\'e}l{\'e}my},\ and\
  \citenamefont {Cazayous}}]{rovillain2010electric}%
  \BibitemOpen
  \bibfield  {author} {\bibinfo {author} {\bibfnamefont {P.}~\bibnamefont
  {Rovillain}}, \bibinfo {author} {\bibfnamefont {R.}~\bibnamefont {De~Sousa}},
  \bibinfo {author} {\bibfnamefont {Y.~T.}\ \bibnamefont {Gallais}}, \bibinfo
  {author} {\bibfnamefont {A.}~\bibnamefont {Sacuto}}, \bibinfo {author}
  {\bibfnamefont {M.~A.}\ \bibnamefont {M{\'e}asson}}, \bibinfo {author}
  {\bibfnamefont {D.}~\bibnamefont {Colson}}, \bibinfo {author} {\bibfnamefont
  {A.}~\bibnamefont {Forget}}, \bibinfo {author} {\bibfnamefont
  {M.}~\bibnamefont {Bibes}}, \bibinfo {author} {\bibfnamefont
  {A.}~\bibnamefont {Barth{\'e}l{\'e}my}}, \ and\ \bibinfo {author}
  {\bibfnamefont {M.}~\bibnamefont {Cazayous}},\ }\href {\doibase
  10.1038/nmat2869} {\bibfield  {journal} {\bibinfo  {journal} {Nat. Mater.}\
  }\textbf {\bibinfo {volume} {9}},\ \bibinfo {pages} {975} (\bibinfo {year}
  {2010})}\BibitemShut {NoStop}%
\bibitem [{\citenamefont {Heron}\ \emph {et~al.}(2014)\citenamefont {Heron},
  \citenamefont {Bosse}, \citenamefont {He}, \citenamefont {Gao}, \citenamefont
  {Trassin}, \citenamefont {Ye}, \citenamefont {Clarkson}, \citenamefont
  {Wang}, \citenamefont {Liu}, \citenamefont {Salahuddin} \emph
  {et~al.}}]{heron2014deterministic}%
  \BibitemOpen
  \bibfield  {author} {\bibinfo {author} {\bibfnamefont {J.~T.}\ \bibnamefont
  {Heron}}, \bibinfo {author} {\bibfnamefont {J.~L.}\ \bibnamefont {Bosse}},
  \bibinfo {author} {\bibfnamefont {Q.}~\bibnamefont {He}}, \bibinfo {author}
  {\bibfnamefont {Y.}~\bibnamefont {Gao}}, \bibinfo {author} {\bibfnamefont
  {M.}~\bibnamefont {Trassin}}, \bibinfo {author} {\bibfnamefont
  {L.}~\bibnamefont {Ye}}, \bibinfo {author} {\bibfnamefont {J.~D.}\
  \bibnamefont {Clarkson}}, \bibinfo {author} {\bibfnamefont {C.}~\bibnamefont
  {Wang}}, \bibinfo {author} {\bibfnamefont {J.}~\bibnamefont {Liu}}, \bibinfo
  {author} {\bibfnamefont {S.}~\bibnamefont {Salahuddin}},  \emph {et~al.},\
  }\href {\doibase 10.1038/nature14010} {\bibfield  {journal} {\bibinfo
  {journal} {Nature}\ }\textbf {\bibinfo {volume} {516}},\ \bibinfo {pages}
  {370} (\bibinfo {year} {2014})}\BibitemShut {NoStop}%
\bibitem [{\citenamefont {Gao}\ \emph {et~al.}(2020)\citenamefont {Gao},
  \citenamefont {Zhang},\ and\ \citenamefont {Xiao}}]{gao2020tunable}%
  \BibitemOpen
  \bibfield  {author} {\bibinfo {author} {\bibfnamefont {Y.}~\bibnamefont
  {Gao}}, \bibinfo {author} {\bibfnamefont {Y.}~\bibnamefont {Zhang}}, \ and\
  \bibinfo {author} {\bibfnamefont {D.}~\bibnamefont {Xiao}},\ }\href {\doibase
  10.1103/PhysRevLett.124.077401} {\bibfield  {journal} {\bibinfo  {journal}
  {Phys. Rev. Lett.}\ }\textbf {\bibinfo {volume} {124}},\ \bibinfo {pages}
  {077401} (\bibinfo {year} {2020})}\BibitemShut {NoStop}%
\bibitem [{\citenamefont {Chen}\ \emph {et~al.}(2024)\citenamefont {Chen},
  \citenamefont {Zhai}, \citenamefont {Xiao},\ and\ \citenamefont
  {Yao}}]{chen2024crossed}%
  \BibitemOpen
  \bibfield  {author} {\bibinfo {author} {\bibfnamefont {C.}~\bibnamefont
  {Chen}}, \bibinfo {author} {\bibfnamefont {D.}~\bibnamefont {Zhai}}, \bibinfo
  {author} {\bibfnamefont {C.}~\bibnamefont {Xiao}}, \ and\ \bibinfo {author}
  {\bibfnamefont {W.}~\bibnamefont {Yao}},\ }\href {\doibase
  10.1103/PhysRevResearch.6.L012059} {\bibfield  {journal} {\bibinfo  {journal}
  {Phys. Rev. Res.}\ }\textbf {\bibinfo {volume} {6}},\ \bibinfo {pages}
  {L012059} (\bibinfo {year} {2024})}\BibitemShut {NoStop}%
\bibitem [{\citenamefont {Fan}\ \emph {et~al.}(2024)\citenamefont {Fan},
  \citenamefont {Xiao},\ and\ \citenamefont {Yao}}]{fan2024intrinsic}%
  \BibitemOpen
  \bibfield  {author} {\bibinfo {author} {\bibfnamefont {F.-R.}\ \bibnamefont
  {Fan}}, \bibinfo {author} {\bibfnamefont {C.}~\bibnamefont {Xiao}}, \ and\
  \bibinfo {author} {\bibfnamefont {W.}~\bibnamefont {Yao}},\ }\href {\doibase
  10.1038/s41467-024-52314-5} {\bibfield  {journal} {\bibinfo  {journal} {Nat.
  Commun.}\ }\textbf {\bibinfo {volume} {15}},\ \bibinfo {pages} {7997}
  (\bibinfo {year} {2024})}\BibitemShut {NoStop}%
\bibitem [{\citenamefont {Li}\ \emph {et~al.}(2024)\citenamefont {Li},
  \citenamefont {Zhai}, \citenamefont {Xiao},\ and\ \citenamefont
  {Yao}}]{li2024dynamical}%
  \BibitemOpen
  \bibfield  {author} {\bibinfo {author} {\bibfnamefont {J.}~\bibnamefont
  {Li}}, \bibinfo {author} {\bibfnamefont {D.}~\bibnamefont {Zhai}}, \bibinfo
  {author} {\bibfnamefont {C.}~\bibnamefont {Xiao}}, \ and\ \bibinfo {author}
  {\bibfnamefont {W.}~\bibnamefont {Yao}},\ }\href {\doibase
  10.1007/s44214-024-00059-z} {\bibfield  {journal} {\bibinfo  {journal}
  {Quantum Frontiers}\ }\textbf {\bibinfo {volume} {3}},\ \bibinfo {pages} {11}
  (\bibinfo {year} {2024})}\BibitemShut {NoStop}%
\bibitem [{\citenamefont {Ohno}\ \emph {et~al.}(2000)\citenamefont {Ohno},
  \citenamefont {Chiba}, \citenamefont {Matsukura}, \citenamefont {Omiya},
  \citenamefont {Abe}, \citenamefont {Dietl}, \citenamefont {Ohno},\ and\
  \citenamefont {Ohtani}}]{ohno2000electric}%
  \BibitemOpen
  \bibfield  {author} {\bibinfo {author} {\bibfnamefont {H.}~\bibnamefont
  {Ohno}}, \bibinfo {author} {\bibfnamefont {A.~D.}\ \bibnamefont {Chiba}},
  \bibinfo {author} {\bibfnamefont {A.~F.}\ \bibnamefont {Matsukura}}, \bibinfo
  {author} {\bibfnamefont {T.}~\bibnamefont {Omiya}}, \bibinfo {author}
  {\bibfnamefont {E.}~\bibnamefont {Abe}}, \bibinfo {author} {\bibfnamefont
  {T.}~\bibnamefont {Dietl}}, \bibinfo {author} {\bibfnamefont
  {Y.}~\bibnamefont {Ohno}}, \ and\ \bibinfo {author} {\bibfnamefont
  {K.}~\bibnamefont {Ohtani}},\ }\href {\doibase 10.1038/35048537} {\bibfield
  {journal} {\bibinfo  {journal} {Nature}\ }\textbf {\bibinfo {volume} {408}},\
  \bibinfo {pages} {944} (\bibinfo {year} {2000})}\BibitemShut {NoStop}%
\bibitem [{\citenamefont {Fert}\ \emph {et~al.}(2024)\citenamefont {Fert},
  \citenamefont {Ramesh}, \citenamefont {Garcia}, \citenamefont {Casanova},\
  and\ \citenamefont {Bibes}}]{fert2024electrical}%
  \BibitemOpen
  \bibfield  {author} {\bibinfo {author} {\bibfnamefont {A.}~\bibnamefont
  {Fert}}, \bibinfo {author} {\bibfnamefont {R.}~\bibnamefont {Ramesh}},
  \bibinfo {author} {\bibfnamefont {V.}~\bibnamefont {Garcia}}, \bibinfo
  {author} {\bibfnamefont {F.}~\bibnamefont {Casanova}}, \ and\ \bibinfo
  {author} {\bibfnamefont {M.}~\bibnamefont {Bibes}},\ }\href {\doibase
  10.1103/RevModPhys.96.015005} {\bibfield  {journal} {\bibinfo  {journal}
  {Rev. Mod. Phys.}\ }\textbf {\bibinfo {volume} {96}},\ \bibinfo {pages}
  {015005} (\bibinfo {year} {2024})}\BibitemShut {NoStop}%
\bibitem [{\citenamefont {Wang}\ \emph {et~al.}(2026)\citenamefont {Wang},
  \citenamefont {Yu}, \citenamefont {Cui}, \citenamefont {Han}, \citenamefont
  {He}, \citenamefont {Wu}, \citenamefont {Zhang}, \citenamefont {Yang},\ and\
  \citenamefont {Yao}}]{wang2026type}%
  \BibitemOpen
  \bibfield  {author} {\bibinfo {author} {\bibfnamefont {Y.}~\bibnamefont
  {Wang}}, \bibinfo {author} {\bibfnamefont {Z.-M.}\ \bibnamefont {Yu}},
  \bibinfo {author} {\bibfnamefont {C.}~\bibnamefont {Cui}}, \bibinfo {author}
  {\bibfnamefont {Y.}~\bibnamefont {Han}}, \bibinfo {author} {\bibfnamefont
  {T.}~\bibnamefont {He}}, \bibinfo {author} {\bibfnamefont {W.}~\bibnamefont
  {Wu}}, \bibinfo {author} {\bibfnamefont {R.-W.}\ \bibnamefont {Zhang}},
  \bibinfo {author} {\bibfnamefont {S.~A.}\ \bibnamefont {Yang}}, \ and\
  \bibinfo {author} {\bibfnamefont {Y.}~\bibnamefont {Yao}},\ }\href {\doibase
  10.1103/PhysRevLett.136.106402} {\bibfield  {journal} {\bibinfo  {journal}
  {Phys. Rev. Lett.}\ }\textbf {\bibinfo {volume} {136}},\ \bibinfo {pages}
  {106402} (\bibinfo {year} {2026})}\BibitemShut {NoStop}%
\bibitem [{sup()}]{supp_mat}%
  \BibitemOpen
  \href@noop {} {\ }\bibinfo {note} {See Supplemental Material for calculation
  methods, the EEE of the nonmagnet FeSe and the altermagnet V2Te2O, and T-even
  and T-odd tensor forms for the conventional EE and EEE.}\BibitemShut {Stop}%
\bibitem [{\citenamefont {Dong}\ \emph {et~al.}(2020)\citenamefont {Dong},
  \citenamefont {Xiao}, \citenamefont {Xiong},\ and\ \citenamefont
  {Niu}}]{dong2020berry}%
  \BibitemOpen
  \bibfield  {author} {\bibinfo {author} {\bibfnamefont {L.}~\bibnamefont
  {Dong}}, \bibinfo {author} {\bibfnamefont {C.}~\bibnamefont {Xiao}}, \bibinfo
  {author} {\bibfnamefont {B.}~\bibnamefont {Xiong}}, \ and\ \bibinfo {author}
  {\bibfnamefont {Q.}~\bibnamefont {Niu}},\ }\href {\doibase
  10.1103/PhysRevLett.124.066601} {\bibfield  {journal} {\bibinfo  {journal}
  {Phys. Rev. Lett.}\ }\textbf {\bibinfo {volume} {124}},\ \bibinfo {pages}
  {066601} (\bibinfo {year} {2020})}\BibitemShut {NoStop}%
\bibitem [{\citenamefont {Garate}\ and\ \citenamefont
  {MacDonald}(2009)}]{garate2009influence}%
  \BibitemOpen
  \bibfield  {author} {\bibinfo {author} {\bibfnamefont {I.}~\bibnamefont
  {Garate}}\ and\ \bibinfo {author} {\bibfnamefont {A.~H.}\ \bibnamefont
  {MacDonald}},\ }\href {\doibase 10.1103/PhysRevB.80.134403} {\bibfield
  {journal} {\bibinfo  {journal} {Phys. Rev. B}\ }\textbf {\bibinfo {volume}
  {80}},\ \bibinfo {pages} {134403} (\bibinfo {year} {2009})}\BibitemShut
  {NoStop}%
\bibitem [{\citenamefont {O'Hara}\ \emph {et~al.}(2018)\citenamefont {O'Hara},
  \citenamefont {Zhu}, \citenamefont {Trout}, \citenamefont {Ahmed},
  \citenamefont {Luo}, \citenamefont {Lee}, \citenamefont {Brenner},
  \citenamefont {Rajan}, \citenamefont {Gupta}, \citenamefont {McComb} \emph
  {et~al.}}]{o2018room}%
  \BibitemOpen
  \bibfield  {author} {\bibinfo {author} {\bibfnamefont {D.~J.}\ \bibnamefont
  {O'Hara}}, \bibinfo {author} {\bibfnamefont {T.}~\bibnamefont {Zhu}},
  \bibinfo {author} {\bibfnamefont {A.~H.}\ \bibnamefont {Trout}}, \bibinfo
  {author} {\bibfnamefont {A.~S.}\ \bibnamefont {Ahmed}}, \bibinfo {author}
  {\bibfnamefont {Y.~K.}\ \bibnamefont {Luo}}, \bibinfo {author} {\bibfnamefont
  {C.~H.}\ \bibnamefont {Lee}}, \bibinfo {author} {\bibfnamefont {M.~R.}\
  \bibnamefont {Brenner}}, \bibinfo {author} {\bibfnamefont {S.}~\bibnamefont
  {Rajan}}, \bibinfo {author} {\bibfnamefont {J.~A.}\ \bibnamefont {Gupta}},
  \bibinfo {author} {\bibfnamefont {D.~W.}\ \bibnamefont {McComb}},  \emph
  {et~al.},\ }\href {\doibase 10.1021/acs.nanolett.8b00608} {\bibfield
  {journal} {\bibinfo  {journal} {Nano Lett.}\ }\textbf {\bibinfo {volume}
  {18}},\ \bibinfo {pages} {3125} (\bibinfo {year} {2018})}\BibitemShut
  {NoStop}%
\bibitem [{\citenamefont {Vanherck}\ \emph {et~al.}(2020)\citenamefont
  {Vanherck}, \citenamefont {Bacaksiz}, \citenamefont {Sor{\'e}e},
  \citenamefont {Milo{\v{s}}evi{\'c}},\ and\ \citenamefont
  {Magnus}}]{vanherck20202d}%
  \BibitemOpen
  \bibfield  {author} {\bibinfo {author} {\bibfnamefont {J.}~\bibnamefont
  {Vanherck}}, \bibinfo {author} {\bibfnamefont {C.}~\bibnamefont {Bacaksiz}},
  \bibinfo {author} {\bibfnamefont {B.}~\bibnamefont {Sor{\'e}e}}, \bibinfo
  {author} {\bibfnamefont {M.~V.}\ \bibnamefont {Milo{\v{s}}evi{\'c}}}, \ and\
  \bibinfo {author} {\bibfnamefont {W.}~\bibnamefont {Magnus}},\ }\href
  {\doibase 10.1063/5.0013302} {\bibfield  {journal} {\bibinfo  {journal}
  {Appl. Phys. Lett.}\ }\textbf {\bibinfo {volume} {117}},\ \bibinfo {pages}
  {052402} (\bibinfo {year} {2020})}\BibitemShut {NoStop}%
\bibitem [{\citenamefont {Ma}\ \emph {et~al.}(2019)\citenamefont {Ma},
  \citenamefont {Xu}, \citenamefont {Shen}, \citenamefont {MacNeill},
  \citenamefont {Fatemi}, \citenamefont {Chang}, \citenamefont {Mier~Valdivia},
  \citenamefont {Wu}, \citenamefont {Du}, \citenamefont {Hsu}, \citenamefont
  {Fang}, \citenamefont {Gibson}, \citenamefont {Watanabe}, \citenamefont
  {Taniguchi}, \citenamefont {Cava}, \citenamefont {Castro}, \citenamefont
  {Lin}, \citenamefont {MacDonald}, \citenamefont {Gedik},\ and\ \citenamefont
  {Jarillo-Herrero}}]{ma2019observation}%
  \BibitemOpen
  \bibfield  {author} {\bibinfo {author} {\bibfnamefont {Q.}~\bibnamefont
  {Ma}}, \bibinfo {author} {\bibfnamefont {S.-Y.}\ \bibnamefont {Xu}}, \bibinfo
  {author} {\bibfnamefont {H.}~\bibnamefont {Shen}}, \bibinfo {author}
  {\bibfnamefont {D.}~\bibnamefont {MacNeill}}, \bibinfo {author}
  {\bibfnamefont {V.}~\bibnamefont {Fatemi}}, \bibinfo {author} {\bibfnamefont
  {T.-R.}\ \bibnamefont {Chang}}, \bibinfo {author} {\bibfnamefont {A.~M.}\
  \bibnamefont {Mier~Valdivia}}, \bibinfo {author} {\bibfnamefont
  {S.}~\bibnamefont {Wu}}, \bibinfo {author} {\bibfnamefont {Z.}~\bibnamefont
  {Du}}, \bibinfo {author} {\bibfnamefont {C.-H.}\ \bibnamefont {Hsu}},
  \bibinfo {author} {\bibfnamefont {S.}~\bibnamefont {Fang}}, \bibinfo {author}
  {\bibfnamefont {Q.~D.}\ \bibnamefont {Gibson}}, \bibinfo {author}
  {\bibfnamefont {K.}~\bibnamefont {Watanabe}}, \bibinfo {author}
  {\bibfnamefont {T.}~\bibnamefont {Taniguchi}}, \bibinfo {author}
  {\bibfnamefont {R.~J.}\ \bibnamefont {Cava}}, \bibinfo {author}
  {\bibfnamefont {E.~V.}\ \bibnamefont {Castro}}, \bibinfo {author}
  {\bibfnamefont {H.}~\bibnamefont {Lin}}, \bibinfo {author} {\bibfnamefont
  {A.}~\bibnamefont {MacDonald}}, \bibinfo {author} {\bibfnamefont
  {N.}~\bibnamefont {Gedik}}, \ and\ \bibinfo {author} {\bibfnamefont
  {P.}~\bibnamefont {Jarillo-Herrero}},\ }\href {\doibase
  10.1038/s41586-018-0807-6} {\bibfield  {journal} {\bibinfo  {journal}
  {Nature}\ }\textbf {\bibinfo {volume} {565}},\ \bibinfo {pages} {337}
  (\bibinfo {year} {2019})}\BibitemShut {NoStop}%
\bibitem [{\citenamefont {Chernyshov}\ \emph {et~al.}(2009)\citenamefont
  {Chernyshov}, \citenamefont {Overby}, \citenamefont {Liu}, \citenamefont
  {Furdyna}, \citenamefont {Lyanda-Geller},\ and\ \citenamefont
  {Rokhinson}}]{chernyshov2009evidence}%
  \BibitemOpen
  \bibfield  {author} {\bibinfo {author} {\bibfnamefont {A.}~\bibnamefont
  {Chernyshov}}, \bibinfo {author} {\bibfnamefont {M.}~\bibnamefont {Overby}},
  \bibinfo {author} {\bibfnamefont {X.}~\bibnamefont {Liu}}, \bibinfo {author}
  {\bibfnamefont {J.~K.}\ \bibnamefont {Furdyna}}, \bibinfo {author}
  {\bibfnamefont {Y.}~\bibnamefont {Lyanda-Geller}}, \ and\ \bibinfo {author}
  {\bibfnamefont {L.~P.}\ \bibnamefont {Rokhinson}},\ }\href {\doibase
  10.1038/nphys1362} {\bibfield  {journal} {\bibinfo  {journal} {Nat. Phys.}\
  }\textbf {\bibinfo {volume} {5}},\ \bibinfo {pages} {656} (\bibinfo {year}
  {2009})}\BibitemShut {NoStop}%
\bibitem [{\citenamefont {Olejn{\'\i}k}\ \emph {et~al.}(2018)\citenamefont
  {Olejn{\'\i}k}, \citenamefont {Seifert}, \citenamefont {Ka{\v{s}}par},
  \citenamefont {Nov{\'a}k}, \citenamefont {Wadley}, \citenamefont {Campion},
  \citenamefont {Baumgartner}, \citenamefont {Gambardella}, \citenamefont
  {N{\v{e}}mec}, \citenamefont {Wunderlich} \emph
  {et~al.}}]{olejnik2018terahertz}%
  \BibitemOpen
  \bibfield  {author} {\bibinfo {author} {\bibfnamefont {K.}~\bibnamefont
  {Olejn{\'\i}k}}, \bibinfo {author} {\bibfnamefont {T.}~\bibnamefont
  {Seifert}}, \bibinfo {author} {\bibfnamefont {Z.}~\bibnamefont
  {Ka{\v{s}}par}}, \bibinfo {author} {\bibfnamefont {V.}~\bibnamefont
  {Nov{\'a}k}}, \bibinfo {author} {\bibfnamefont {P.}~\bibnamefont {Wadley}},
  \bibinfo {author} {\bibfnamefont {R.~P.}\ \bibnamefont {Campion}}, \bibinfo
  {author} {\bibfnamefont {M.}~\bibnamefont {Baumgartner}}, \bibinfo {author}
  {\bibfnamefont {P.}~\bibnamefont {Gambardella}}, \bibinfo {author}
  {\bibfnamefont {P.}~\bibnamefont {N{\v{e}}mec}}, \bibinfo {author}
  {\bibfnamefont {J.}~\bibnamefont {Wunderlich}},  \emph {et~al.},\ }\href
  {\doibase 10.1126/sciadv.aar3566} {\bibfield  {journal} {\bibinfo  {journal}
  {Sci. Adv.}\ }\textbf {\bibinfo {volume} {4}},\ \bibinfo {pages} {eaar3566}
  (\bibinfo {year} {2018})}\BibitemShut {NoStop}%
\bibitem [{\citenamefont {Zhou}\ \emph {et~al.}(2019)\citenamefont {Zhou},
  \citenamefont {Chen}, \citenamefont {Zhang}, \citenamefont {Li},
  \citenamefont {Shi}, \citenamefont {Sun}, \citenamefont {Saleem},
  \citenamefont {You}, \citenamefont {Pan},\ and\ \citenamefont
  {Song}}]{zhou2019fieldlike}%
  \BibitemOpen
  \bibfield  {author} {\bibinfo {author} {\bibfnamefont {X.~F.}\ \bibnamefont
  {Zhou}}, \bibinfo {author} {\bibfnamefont {X.~Z.}\ \bibnamefont {Chen}},
  \bibinfo {author} {\bibfnamefont {J.}~\bibnamefont {Zhang}}, \bibinfo
  {author} {\bibfnamefont {F.}~\bibnamefont {Li}}, \bibinfo {author}
  {\bibfnamefont {G.~Y.}\ \bibnamefont {Shi}}, \bibinfo {author} {\bibfnamefont
  {Y.~M.}\ \bibnamefont {Sun}}, \bibinfo {author} {\bibfnamefont {M.~S.}\
  \bibnamefont {Saleem}}, \bibinfo {author} {\bibfnamefont {Y.~F.}\
  \bibnamefont {You}}, \bibinfo {author} {\bibfnamefont {F.}~\bibnamefont
  {Pan}}, \ and\ \bibinfo {author} {\bibfnamefont {C.}~\bibnamefont {Song}},\
  }\href {\doibase 10.1103/PhysRevApplied.11.054030} {\bibfield  {journal}
  {\bibinfo  {journal} {Phys. Rev. Appl.}\ }\textbf {\bibinfo {volume} {11}},\
  \bibinfo {pages} {054030} (\bibinfo {year} {2019})}\BibitemShut {NoStop}%
\bibitem [{\citenamefont {Kurebayashi}\ \emph {et~al.}(2014)\citenamefont
  {Kurebayashi}, \citenamefont {Sinova}, \citenamefont {Fang}, \citenamefont
  {Irvine}, \citenamefont {Skinner}, \citenamefont {Wunderlich}, \citenamefont
  {Nov{\'a}k}, \citenamefont {Campion}, \citenamefont {Gallagher},
  \citenamefont {Vehstedt} \emph {et~al.}}]{kurebayashi2014antidamping}%
  \BibitemOpen
  \bibfield  {author} {\bibinfo {author} {\bibfnamefont {H.}~\bibnamefont
  {Kurebayashi}}, \bibinfo {author} {\bibfnamefont {J.}~\bibnamefont {Sinova}},
  \bibinfo {author} {\bibfnamefont {D.}~\bibnamefont {Fang}}, \bibinfo {author}
  {\bibfnamefont {A.~C.}\ \bibnamefont {Irvine}}, \bibinfo {author}
  {\bibfnamefont {T.~D.}\ \bibnamefont {Skinner}}, \bibinfo {author}
  {\bibfnamefont {J.}~\bibnamefont {Wunderlich}}, \bibinfo {author}
  {\bibfnamefont {V.}~\bibnamefont {Nov{\'a}k}}, \bibinfo {author}
  {\bibfnamefont {R.~P.}\ \bibnamefont {Campion}}, \bibinfo {author}
  {\bibfnamefont {B.~L.}\ \bibnamefont {Gallagher}}, \bibinfo {author}
  {\bibfnamefont {E.~K.}\ \bibnamefont {Vehstedt}},  \emph {et~al.},\ }\href
  {\doibase 10.1038/nnano.2014.15} {\bibfield  {journal} {\bibinfo  {journal}
  {Nat. Nanotechnol.}\ }\textbf {\bibinfo {volume} {9}},\ \bibinfo {pages}
  {211} (\bibinfo {year} {2014})}\BibitemShut {NoStop}%
\bibitem [{\citenamefont {Shi}\ and\ \citenamefont
  {Kioupakis}(2015)}]{shi2015anisotropic}%
  \BibitemOpen
  \bibfield  {author} {\bibinfo {author} {\bibfnamefont {G.}~\bibnamefont
  {Shi}}\ and\ \bibinfo {author} {\bibfnamefont {E.}~\bibnamefont
  {Kioupakis}},\ }\href {\doibase 10.1021/acs.nanolett.5b02861} {\bibfield
  {journal} {\bibinfo  {journal} {Nano Lett.}\ }\textbf {\bibinfo {volume}
  {15}},\ \bibinfo {pages} {6926} (\bibinfo {year} {2015})}\BibitemShut
  {NoStop}%
\bibitem [{\citenamefont {Li}\ \emph {et~al.}(2013)\citenamefont {Li},
  \citenamefont {Chen}, \citenamefont {Hu}, \citenamefont {Wang}, \citenamefont
  {Zhang}, \citenamefont {Chen},\ and\ \citenamefont {Wang}}]{li2013single}%
  \BibitemOpen
  \bibfield  {author} {\bibinfo {author} {\bibfnamefont {L.}~\bibnamefont
  {Li}}, \bibinfo {author} {\bibfnamefont {Z.}~\bibnamefont {Chen}}, \bibinfo
  {author} {\bibfnamefont {Y.}~\bibnamefont {Hu}}, \bibinfo {author}
  {\bibfnamefont {X.}~\bibnamefont {Wang}}, \bibinfo {author} {\bibfnamefont
  {T.}~\bibnamefont {Zhang}}, \bibinfo {author} {\bibfnamefont
  {W.}~\bibnamefont {Chen}}, \ and\ \bibinfo {author} {\bibfnamefont
  {Q.}~\bibnamefont {Wang}},\ }\href {\doibase 10.1021/ja3108017} {\bibfield
  {journal} {\bibinfo  {journal} {J. Am. Chem. Soc.}\ }\textbf {\bibinfo
  {volume} {135}},\ \bibinfo {pages} {1213} (\bibinfo {year}
  {2013})}\BibitemShut {NoStop}%
\bibitem [{\citenamefont {Fang}\ \emph {et~al.}(2016)\citenamefont {Fang},
  \citenamefont {Zhang}, \citenamefont {Qin}, \citenamefont {Yan},
  \citenamefont {Zheng},\ and\ \citenamefont {Su}}]{fang2016layer}%
  \BibitemOpen
  \bibfield  {author} {\bibinfo {author} {\bibfnamefont {W.}~\bibnamefont
  {Fang}}, \bibinfo {author} {\bibfnamefont {L.-C.}\ \bibnamefont {Zhang}},
  \bibinfo {author} {\bibfnamefont {G.}~\bibnamefont {Qin}}, \bibinfo {author}
  {\bibfnamefont {Q.-B.}\ \bibnamefont {Yan}}, \bibinfo {author} {\bibfnamefont
  {Q.-R.}\ \bibnamefont {Zheng}}, \ and\ \bibinfo {author} {\bibfnamefont
  {G.}~\bibnamefont {Su}},\ }\href {\doibase 10.48550/arXiv.1603.01791}
  {\bibfield  {journal} {\bibinfo  {journal} {arXiv preprint}\ } (\bibinfo
  {year} {2016}),\ 10.48550/arXiv.1603.01791}\BibitemShut {NoStop}%
\end{thebibliography}%

\end{document}